%% file: lmcrev1.tex
\documentstyle[amssymb,psfig]{mn}

\title[Mass segregation in young LMC clusters I.]{Mass segregation in
young compact star clusters in the Large Magellanic Cloud: I.  Data and
Luminosity Functions}

\author[R.  de Grijs et al.]{R.  de Grijs\thanks{E-mail:
grijs@ast.cam.ac.uk}, R.A.  Johnson\thanks{Present address: European
Southern Observatory, Alonso de Cordova 3107, Santiago 19, Chile}, G.F. 
Gilmore and C.M.  Frayn \\
Institute of Astronomy, University of Cambridge, Madingley Road,
Cambridge CB3 0HA
}

\date{Accepted ---. Received ---; in original form ---.}

\pubyear{2001}

\begin{document}
\maketitle

\begin{abstract}
We have undertaken a detailed analysis of {\sl HST/WFPC2} and STIS
imaging observations, and of supplementary wide-field ground-based
observations obtained with the NTT of two young ($\sim 10-25$ Myr)
compact star clusters in the LMC, NGC 1805 and NGC 1818.  The ultimate
goal of our work is to improve our understanding of the degree of
primordial mass segregation in star clusters. This is crucial for the
interpretation of observational luminosity functions (LFs) in terms of
the initial mass function (IMF), and for constraining the universality
of the IMF. \\
We present evidence for strong luminosity segregation in both clusters. 
The LF slopes steepen with cluster radius; in both NGC 1805 and NGC 1818
the LF slopes reach a stable level well beyond the clusters' core or
half-light radii.  In addition, the brightest cluster stars are strongly
concentrated within the inner $\sim 4 R_{\rm hl}$. \\
The global cluster LF, although strongly nonlinear, is fairly well
approximated by the core or half-light LF; the (annular) LFs at these
radii are dominated by the segregated high-luminosity stars, however. \\
We present tentative evidence for the presence of an excess number of
bright stars surrounding NGC 1818, for which we argue that they are most
likely massive stars that have been collisionally ejected from the
cluster core.  We therefore suggest that the cores of massive young
stars clusters undergo significant dynamical evolution, even on
time-scales as short as $\sim 25$ Myr. 
\end{abstract}


\section{Introduction: Mass Segregation and its Implications}
\label{intro.sec}

One of the major uncertainties in modern astrophysics is the issue
of whether the stellar initial mass function (IMF) is universal or,
alternatively, determined by environmental effects.  Galactic globular
clusters (GCs) and rich, compact Magellanic Cloud star clusters are
ideal laboratories for providing strong constraints on the universality
of the IMF, in particular because they are essentially single age,
single metallicity systems for which statistically significant samples
of individual stars over a range of masses can easily be resolved. 

However, the effects of mass segregation in both young and old star
clusters, in the sense that the more massive stars are more centrally
concentrated than the lower-mass stars, clearly complicate the
interpretation of an observed luminosity function (LF) at a given
position within a star cluster in terms of its IMF.  Without reliable
corrections for the effects of mass segregation, hence for the structure
and dynamical evolution of the cluster, it is impossible to obtain an
accurate estimate of the IMF from the observational LF. 

In this paper, we will concentrate on the analysis of the behaviour of
the stellar LF of two young Large Magellanic Cloud (LMC) star clusters
as a function of radius; in a companion paper (de Grijs et al.  2001,
Paper II), we will derive the associated mass functions (MFs) and
discuss in detail how this transformation depends on the assumed models. 
Finally, in a forthcoming paper (de Grijs et al., in prep., Paper III),
we will extend our analysis to our larger LMC cluster sample, spanning a
significant age range, in which we will attempt to model the dynamical
cluster evolution. 

Although the standard picture, in which stars in dense clusters evolve
rapidly towards a state of energy equipartition through stellar
encounters, with the corresponding mass segregation, is generally
accepted, observations of various degrees of mass segregation in very
young star clusters (e.g., Hillenbrand 1997, Testi et al.  1997, Fischer
et al.  1998, Hillenbrand \& Hartmann 1998, Hillenbrand \& Carpenter
2000) suggest that at least some of this effect is related to the
process of star and star cluster formation itself (cf.  Paper II): these
clusters are often significantly younger than their two-body relaxation
time (even the equivalent relaxation time in the core), the time scale
on which dynamical mass segregation should occur.  Quantifying the
degree of actual mass segregation is thus crucial for the interpretation
of observational LFs in terms of the IMF, even for very young star
clusters. 

Although dynamical two-body relaxation effects tend to segregate a
cluster's mass distribution roughly on its mean relaxation time scale,
possibly enhanced by some degree of primordial mass segregation, the
observability of mass segregation depends on a number of conditions. 
These are mostly interrelated and include, among others, the cluster's
degree of central concentration, its age, the radial range sampled by
the observations and, most of all, the observer's ability to resolve the
individual cluster stars towards the cluster centre.  {\sl Hubble Space
Telescope (HST)} observations show, almost without exception, mass
segregation effects in old Galactic GCs, in rich, compact Magellanic
Cloud clusters, and in young star clusters still embedded in molecular
clouds, although to varying degrees.  Not surprisingly, the fraction of
ground-based observations reporting the detection of mass segregation
effects, sometimes even in the same objects, is significantly lower.  We
will now briefly discuss the observational evidence for mass segregation
in the youngest star clusters. 

\subsection{Rich Compact Magellanic Cloud Clusters}
\label{lmcclus.sect}

Since the rich, compact star clusters in the Magellanic Clouds span a
wide range of ages, from $\sim 10^6$ to $\sim 1.3 \times 10^{10}$ yr
(cf.  Elson \& Fall 1988), observable effects of mass segregation due to
two-body relaxation need not be expected {\em a priori}, particularly
for the youngest clusters.  Ground-based observations may be severely
hampered by crowding in the cluster cores, thus hiding possible
signatures of dynamical, as well as primordial, mass segregation.  This
is likely the case in a number of the young star clusters in the LMC
studied by Elson, Fall \& Freeman (1987) and Subramaniam, Sagar \& Bhatt
(1993), although the uncertainties in the power-law slopes of the MFs
derived by Elson et al.  (1987), down to a given limiting brightness,
are large and could possibly hide mass segregation signatures. 

The quality of ground-based data for rich LMC clusters often allows only
marginal confirmation of possible mass segregation effects, especially
in young clusters if based on LF differences as a function of cluster
radius (e.g., NGC 1711, Subramaniam et al.  1993; NGC 1866, Elson et al. 
1987).  Similarly, ground-based observations of H4 (Mateo \& Hodge
1986), LW 79 (Mateo \& Hodge 1987), and ESO121-SC09 (Mateo, Hodge \&
Schommer 1986, Papenhausen \& Schommer 1988) report either no convincing
(LW 79) or marginal evidence for mass segregation based on the
distributions of main-sequence and giant-branch stars, due to
small-number statistics in the cluster cores. 

In all other LMC clusters studied for this purpose to date, using either
LFs or their associated MFs as a function of radius to assess the
distributions of stars of varying brightness (mass), ground-based
observations (e.g., NGC 2100, Westerlund 1961; NGC 2098 and SL 666,
Kontizas et al.  1998) do indeed show strong indications of mass
segregation.  In addition, observations with the {\sl HST} have also
resulted in convincing cases for mass segregation, e.g., in all objects
in our own LMC cluster sample (cf.  Elson et al.  1999 for NGC 1868, and
Santiago et al.  2001; hereafter SBJG), in NGC 2157 (Fischer et al. 
1998), and in the Small Magellanic Cloud (SMC) star cluster NGC 330
(Sirianni et al.  2001).  The results of SBJG were based on a first,
preliminary analysis of the {\sl WFPC2} observations also used in this
paper.  Here, we present a more detailed analysis of the effects of
luminosity segregation in the two youngest clusters in our LMC sample,
while combining the {\sl WFPC2} data with the STIS observations also
obtained as part of the same programme (cf.  Section \ref{sample.sec}),
and with wide-field ground-based observations obtained with the NTT. 

We note that for one of the objects in our sample, NGC 1818,
contradicting results have been obtained: Hunter et al.  (1997) did not
find evidence for mass segregation in this cluster for stellar masses in
the range $0.85 \le m \le 9 M_\odot$, with the proviso that the cluster
core does contain brighter stars that were not included in their study
due to saturation, whereas the outer regions do not.  Recently, however,
SBJG confirmed significant mass segregation, based on the radial
variance of the cluster LFs.  Elson et al.  (1998) adopted a different
approach, and concluded that the fraction of binary stars increases
significantly from $\sim 20$\% in the outer regions to $\sim 35$\% in
the core.  This is consistent with expectations from {\it (i)} dynamical
mass segregation, where we might expect the more massive binaries to
have undergone substantial two-body relaxation while single stars would
not have (cf.  Paper II), and {\it (ii)} the radial dependence of the
binary creation and destruction rates in these young star clusters. 
Elson et al.  (1998) show that their derived IMF is fully consistent
with Hunter et al.'s (1997) results. 

\subsection{Very Young Star Clusters}

The study of proto-stellar clusters, i.e., very young associations of
stars still embedded in the molecular clouds from which they originated,
might give us a handle to constrain the degree of primordial mass
segregation.  In the Galaxy, in three such young star clusters (YSCs)
mass segregation effects have been studied in detail in the past decade. 

Ground-based observations of both NGC 2024 (Lada et al.  1991, Carpenter
et al.  1997) and the Monoceros R2 (MonR2) complex (Carpenter et al. 
1997) have not been able to show the presence or absence of (possibly
primordial) mass segregation convincingly.  While Lada et al.  (1991)
suggested that the brighter stars in NGC 2024 seem to be more centrally
concentrated than the fainter cluster members, this evidence was deemed
inconclusive by Carpenter et al.  (1997).  They argued that this result
was based on an incomplete sample of cluster stars, although mass
segregation might be limited to the OB stars forming in the very centre. 
These same authors argued that for masses below $2 M_\odot$, mass
segregation effects in MonR2 amount to only a $\sim 2 \sigma$ result,
although the most massive star ($\sim 10 M_\odot$) appears to be forming
near the cluster centre, where their extinction-limited sample of
cluster stars does not reach. 

A combination of both ground-based (e.g., Hillenbrand 1997) and {\sl
HST} observations (e.g., Hillenbrand \& Hartmann 1998) of the Orion
Nebula Cluster (ONC), and in particular of its very core, the Trapezium
stars, have presented clear evidence for mass segregation for the $m > 5
M_\odot$ component, with some evidence for general mass segregation down
to $m \simeq 1$--$2 M_\odot$ (Hillenbrand \& Hartmann 1998, see also the
review by Larson 1993).  Mass segregation in this YSC has, in fact, been
known for more than 5 decades, as pointed out by Hillenbrand \& Hartmann
(1998). 

Finally, R136, the central cluster in the actively star-forming 30
Doradus region in the LMC (age $\lesssim$ 3--4 Myr, cf.  Hunter et al. 
1995), has been studied extensively, both from the ground and with {\sl
HST}.  A variety of techniques have revealed a significant overabundance
of high-mass stars in its very centre, and thus strong mass segregation
(e.g., Campbell et al.  1992, Larson 1993, Malumuth \& Heap 1994, Brandl
et al.  1996).  Hunter et al.  (1995) constrained the central region in
which possible mass segregation effects were observed to $r \lesssim
0.5$ pc, and found little evidence of mass segregation beyond this
radius for stars in the mass range $2.8 \le m \le 15 M_\odot$.  However,
they reported a hint of an excess of bright stars within 0.5 pc, and a
deficit of the highest-mass stars in the annulus $0.6 \le r \le 1.2$ pc,
which would be consistent with expectations for mass segregation. 
Brandl, Chernoff \& Moffat (2001), finally, present very interesting
evidence for the ejection of a fair number of massive stars from the
core, due to two-body encounters, which could not, with higher
confidence, be attributed to alternative scenarios. 

Thus, in most of the young compact star clusters that can be resolved in
individual stars by currently available telescopes, mass segregation
effects are observed, {\em although to varying degrees}.  This
underlines the importance of our understanding of the physical processes
involved in the formation and evolution of star clusters, and in
particular of the IMF, which will ultimately determine whether a young
star cluster will eventually be destroyed or evolve to a Galactic
GC-type object. 

Because of their large range in ages, rich compact star clusters in the
LMC are ideal candidates to assess the effects and magnitude of
dynamical mass segregation, provided that one can constrain the degree
of primordial mass segregation to within reasonable uncertainties. 
Constraining this degree of primordial mass segregation is of crucial
importance to discriminate among different theories of star formation in
clusters, which generally depend on the presence of extensive
dissipation processes.  The presence or absence of dissipation during
the star formation process may have significantly different effects on
the radial dependence of the IMF (cf.  Fischer et al.  1998). 

\section{The LMC Cluster Sample}
\label{sample.sec}

In this paper, we will focus on the two youngest LMC star clusters in
our sample, NGC 1805 and NGC 1818.  Table \ref{sample.tab} contains the
current best estimates available in the literature for a few of the most
important properties of each cluster, including their age, metallicity,
core radii, mass, adopted distance modulus, foreground reddening E({\it
B--V}), and median relaxation time-scale.  For a full overview of the
clusters' physical parameters, we refer the reader to {\tt
http://www.ast.cam.ac.uk/STELLARPOPS/LMCdatabase/}. 

\input{yclusparams.tex}

\subsection{Observations}

To study the effects of mass segregation properly, the observational
data needs to meet the following conditions: 

\begin{enumerate}

\item The observability of the effects of {\em dynamical} mass
segregation is a strong function of a cluster's age (cf.  Paper II) and
the observer's ability to resolve individual stars, in particular in its
core.  Due to crowding, the observations in the core need to be of high
resolution. 

\item Since the initial mass of stars at the main sequence turn-off
magnitude and the mass of stars on the giant and horizontal branches is
approximately constant, we will need to obtain deep observations,
extending well down the main sequence, in order to sample a sufficiently
large mass range.  In fact, as Bolte (1989) argued, since the dynamical
relaxation time scale for the cluster stars, with the exception of those
in the very core, is long compared to the time scale for mass loss
between the main sequence and horizontal branch, we can in principle
extend our analysis to include stars from the horizontal branch down to
the completeness limit of the main sequence. 

\end{enumerate}

As part of {\sl HST} GO programme 7307, we obtained {\sl WFPC2} and STIS
imaging observations of the populous LMC clusters in Table
\ref{sample.tab}.  The high resolution of the {\sl WFPC2}/PC
observations ($\sim 1.8$ pixels; the pixel size of the WF chips is
$0.097''$, with a total combined field of view of roughly 4850
arcsec$^2$ for the entire {\sl WFPC2} detector, while the PC pixel size
is $0.0455''$) meets the first condition above, while the deep STIS data
allow for the construction of very deep LFs down to $\sim 0.2 M_\odot$;
in fact, STIS (in imaging mode through long-pass [LP] filters) is five
times more sensitive for faint red objects than {\sl WFPC2}. 

Although parts of the observations of the LMC clusters have been
described elsewhere already (e.g., Beaulieu et al.  1999, Elson et al. 
1999, Castro et al.  2001, Johnson et al.  2001, SBJG), here we will
give a brief overview of the available data for the two youngest star
clusters in our sample.  This paper builds on the preparatory research
by SBJG, which we will continue in significantly greater detail;
colour-magnitude diagrams (CMDs) of these clusters were published by
Johnson et al.  (2001).  These latter authors focused predominantly on
the brighter stars in NGC 1805 and NGC 1818 to consider ages and age
ranges.  In this study we include the entire range of magnitudes down to
the 50\% completeness limit (Sec.  \ref{compl.sect}) to consider the
evidence for luminosity and mass segregation in these clusters. 

\subsubsection{WFPC2 observations}

\begin{figure*}
\vspace*{-0.3cm}
\psfig{figure=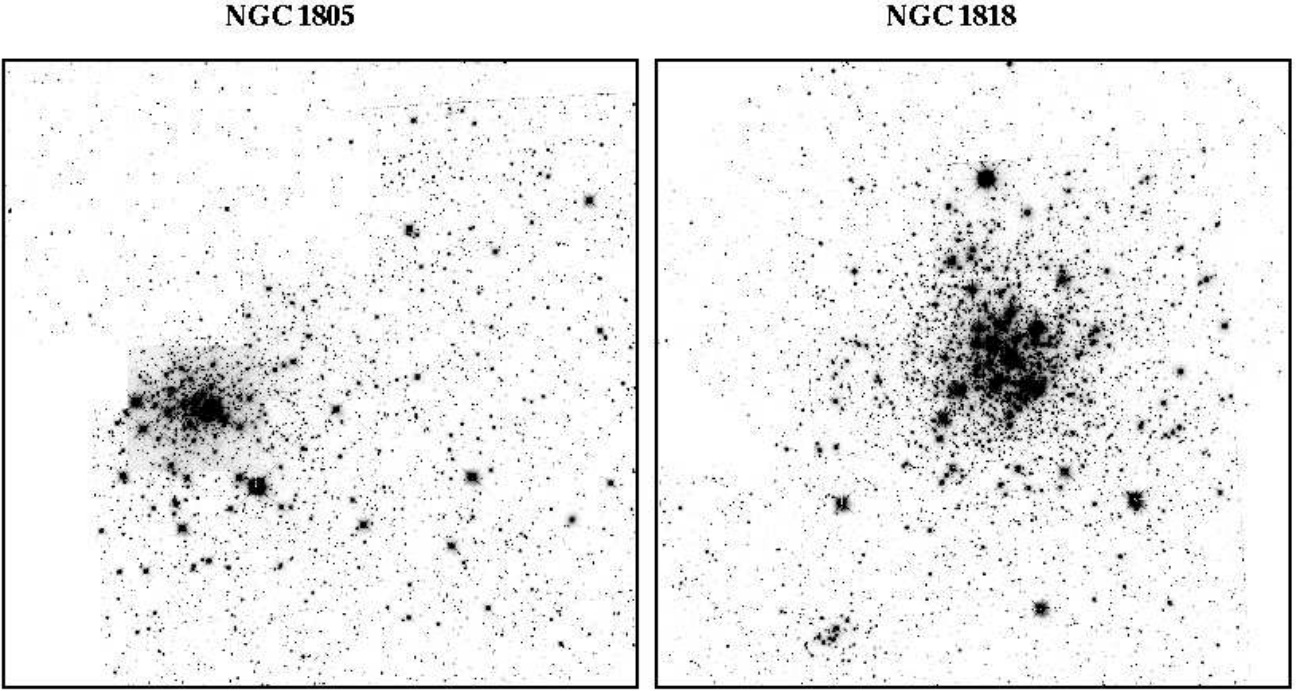,width=18cm}
\vspace*{-0.3cm}
\caption{\label{images.fig}Combined CEN and HALF {\sl WFPC2} F555W
images of NGC 1805 and NGC 1818.  Each panel is $164 \times 164''$
($\sim 41 \times 41$ pc); North is up, East to the left. For reasons of
clarity and dynamic range, the central cluster regions are represented
by the PC fields of view of the 3x140s CEN exposures, since they are
mostly saturated in the HALF fields.}
\end{figure*}

We obtained {\sl WFPC2} exposures through the F555W and F814W filters
(roughly corresponding to the Johnson-Cousins {\it V} and {\it I}
filters, resp.) for each cluster, with the PC centred on both the
cluster centre, and on its half-mass radius.  Following SBJG, we will
refer to these two sets of exposures as our CEN and HALF fields,
respectively.  For the CEN fields, we obtained both deep (exposure times
of 140s and 300s, respectively, for each individual image in F555W and
F814W) and shallow (exposure times of 5s and 20s were used for the F555W
and F814W filters, respectively) images.  The shallow exposures were
intended to obtain aperture photometry for the brightest stars in the
cluster centres, which are saturated in the deeper exposures.  At each
position, for each set of deep and shallow exposures, and through both
filters, we imaged our clusters in sets of 3 individual observations, to
facilitate the removal of cosmic rays (Sec.  \ref{reduction.sect}).  For
the HALF field, we obtained deep observations with a total exposure time
of 2500s through each filter.  The saturation level for all of the
individual calibrated observations, as used in our subsequent analysis,
is roughly 3600 counts.  A summary of our CEN and HALF {\sl WFPC2}
observations is included in Table \ref{obslog.tab}.  In Fig. 
\ref{images.fig} we show the composite (combined CEN and HALF fields)
{\sl WFPC2} F555W images of the two clusters. 

Note the small subcluster $\sim 90''$ (22 pc) South East of NGC 1818;
this is cluster NGC 1818 B (cf.  Will et al.  1995, Grebel 1997),
located at RA(J2000) = 05:04:01.82; Dec (J2000) = $-$66:26:46.1.  NGC
1818 B is found to be of similar age, $\sim 30$ Myr, as NGC 1818 (Grebel
1997), although it is not clear whether NGC 1818 B is associated with
NGC 1818.  The configuration of both clusters is reminiscent of the LMC
cluster pair NGC 1850 and NGC 1850 A, which are apparently physically
associated with each other (cf.  Caloi \& Cassatella 1998). 

\input{obslog.tex}

In addition, parallel {\sl WFPC2} fields, located $\sim 7'$ from each
cluster centre, were obtained with NICMOS as primary detector.  These
fields, with total exposure times (two images combined) of 1200s and
800s for F555W and F814W, respectively, will be used for the subtraction
of the background LMC field stars and unresolved background galaxies
(Sec.  \ref{approach.sec}; cf.  Castro et al.  2001, SBJG).  We will
refer to these fields as ``NGC 1805-par'' and ``NGC 1818-par'',
respectively.  In addition, we also obtained deeper {\sl WFPC2}
observations of a background field (``Background-1''), with exposure
times of 7800s and 5200s in F555W and F814W, respectively. This field
was selected to represent the faint LMC disk background at a similar
distance from the bar to the clusters of interest here.

The {\sl WFPC2} results presented in this paper are based on an
independent reprocessing of the data used by SBJG (using slightly
different methods) as a check that the results are not sensitive to the
data reduction techniques applied. 

\subsubsection{STIS imaging observations}

We also obtained deep STIS CCD observations of small sections of the
cluster HALF fields in ACCUM imaging mode through the F28$\times$50LP
filter.  The fields were exposed for 2950s each in sets of 5
observations (see also Elson et al.  1999; Table \ref{obslog.tab}); each
observation was split into 2 exposures to allow for the removal of
cosmic rays by the data processing pipeline.  The STIS CCD covers a
nominal $52'' \times 52''$ square field of view, with $1024 \times 1024
\sim 0.05''$ pixels. The actual field of view of our combined HALF
fields measures $48 \times 28''$, for both clusters.

\subsubsection{Wide-field ground-based observations}

Finally, we obtained wide-field ground-based observations of both LMC
clusters with the ESO New Technology Telescope (NTT) at La Silla, Chile,
equipped with the EMMI wide field imager.  These observations were
obtained in service mode on 1 September 2000 using the standard {\it V}
(\#606), and {\it I} (\#610) filters under photometric conditions (see
Table \ref{obslog.tab} for the observational characteristics); the
seeing FWHM for these observations ranged roughly between 0.8 and
$1.0''$.  We used EMMI's red Tektronix TK2048 EB Grade 2 CCD, \#36, with
a pixel size of $0.27''$ and a field of view of $9.15' \times 8.6'$. 
The data were reduced following standard reduction procedures;
calibration solutions were obtained by using stars in common between the
NTT and {\sl WFPC2} fields. 

\subsection{Image Processing} 
\label{reduction.sect}

Pipeline image reduction and recalibration of the {\sl WFPC2} and STIS
images was done with standard procedures provided as part of the
IRAF/STSDAS\footnote{The Image Reduction and Analysis Facility (IRAF) is
distributed by the National Optical Astronomy Observatories, which is
operated by the Association of Universities for Research in Astronomy,
Inc., under cooperative agreement with the National Science Foundation. 
STSDAS, the Space Telescope Science Data Analysis System, contains tasks
complementary to the existing IRAF tasks.  We used Version 2.2 (August
2000) for the data reduction performed in this paper.} package, using
the updated and corrected on-orbit flat fields and related reference
files most appropriate for our observations. 

Since our {\sl WFPC2} images with a common pointing were aligned to
within a few hundredths of a pixel, we simply co-added the individual
observations in a given filter for each of the CEN or HALF fields using
the IRAF task {\sc crrej} to improve the signal-to-noise ratio in our
images.  This task also removed cosmic ray events in a series of
iterations that allow correction for cosmic ray hits in pixels adjacent
to those that already have been corrected in an earlier iteration. 
After some experimenting, we found that as few as 3 iterations produce
output images in which the remaining cosmic ray hits -- if any -- are
indistinguishable from poisson noise.  We carefully checked that the
corruption of the centres of the brighter stars in our images by the
cosmic ray rejection routines, due to slight misalignments among the 3
individual exposures, was kept to a minimum by comparing the results of
aperture photometry before and after combining our science exposures. 

Alignment of the individual recalibrated STIS images was ensured by
running IRAF's {\sc imalign} task (requiring shifts of $\lesssim 2$
pixels in either direction), upon which a median-filtered final image
was produced using the {\sc imcombine} task. 

We used the final images for each of the cluster fields and for each
filter to obtain source lists (Sec. \ref{sources.sec}) and to subsequently
perform aperture photometry on these source lists (Sec. \ref{apphot.sec}). 

\subsection{Source Selection}
\label{sources.sec}

We based our initial selection of source candidates on a modified
version of the {\sc daofind} task in the {\sc daophot} software package
(Stetson 1987), running under {\sc idl}.\footnote{The Interactive Data 
Language (IDL) is licensed by Research Systems Inc., of Boulder, CO.}

For both the CEN and HALF fields, we decided to combine and
cross-correlate the source lists obtained in both the F555W and F814W
passbands to define our master source lists, for each cluster field.  We
allowed only a 1-pixel positional mismatch between the individual source
detections in F555W and F814W.  After some experimenting, we decided to
set our detection thresholds at four times the appropriate sky
background noise level ($\sigma_{\rm bg}$) in each of our final {\sl
WFPC2} images.  This turned out to be the best compromise between
including the maximum number of genuine source detections and the
minimum number of spurious features due to noise or artifacts along
diffraction spikes of bright saturated stars in the fields. 

The automated source detection routine, used with suitable constraints
on source sharpness and roundness parameters, and the subsequent
cross-correlation technique resulted in combined source lists largely
devoid of non-stellar objects and saturated stars, thus minimising
potential contamination of our photometry.  However, to prevent
contamination by background galaxies or possibly remaining instrumental
artifacts, we employed a second size selection criterion to our master
source lists, using a Gaussian fitting routine to estimate source
extent.  Although the {\sl WFPC2} PSF is decidedly non-Gaussian, this
technique, when applied uniformly to all objects, provides a fairly
accurate assessment of the relative object sizes, thus allowing us to
distinguish between stars and non-stellar objects.  Measurements of
stars in our images, and comparison with both stars from the {\sl WFPC2}
PSF library and artificial PSFs produced by Tiny Tim (Krist \& Hook
1997), showed that the Gaussian $\sigma$ of the {\sl WFPC2} PSF profile
is of order 0.80--0.85 WF or PC pixels; we therefore only retained
sources with $0.55 \le \sigma_{\rm Gauss} \le 1.15$, thus allowing for
the non-Gaussian PSF shape and for effects of instrumental noise on the
measured sizes, in particular of the fainter objects.  We are very
confident that our final source lists thus obtained exclusively contain
genuine stars. 

For the STIS images, we also employed a 4 $\sigma_{\rm bg}$
detection limit and similar final source list selection techniques. 

The total numbers of genuine stars for each cluster field are listed in
Table \ref{sources.tab}.

\input{sources.tex}

\subsection{Aperture Photometry}
\label{apphot.sec}

Because of the undersampling of point sources by the {\sl WFPC2} optics,
aperture photometry produces more accurate magnitude and colour
measurements than PSF fitting, in not too crowded fields (cf.  Castro et
al.  2001, SBJG).  Therefore, we obtained aperture photometry of the
stars in our final source lists, using 2-pixel aperture radii.  This
corresponds to $0.2''$ radii for the 3 WF chips and $r = 0.09''$ for the
PC.  A 2-pixel aperture radius is close to the optimum radius for
stellar aperture photometry in rich star clusters; it produces the
smallest photometric errors and the narrowest main sequence for our
sample clusters.  It is a compromise between the need to include the
core of the PSF but avoid light contamination by neighbouring objects. 

In addition, to avoid runaway photometric errors due to steep gradients
in the background light because of nearby bright stars or diffraction
spikes, we included a criterion to reject sources with mode(background)
$>$ (background $\, + 25 \times \sigma_{\rm bg}$).  Extensive
experimentation showed that this rejection limit only excluded objects
for which the photometry was genuinely affected by nearby bright or
saturated stars.  The number of stars rejected at this stage is also
listed in Table \ref{sources.tab}.  The relatively large number of
rejected sources in the HALF fields of either cluster is due to fact
that the parts of the cluster centres imaged by these exposures contain
a large number of bright, saturated stars (due to the long exposure
times), in the vicinity of which the photometric errors are unacceptably
large. 

Significant aperture corrections (ACs), determined from the individual
data frames, were then applied to the measured magnitudes.  For the
individual {\sl WFPC2} chips, we used the position-dependent ACs to
$0.5''$ apertures used by Johnson et al.  (2001), defined as AC = $a + b
\, r_c$, where $r_c$ is the distance from the centre of the chip in
pixels.  These ACs were found to be the same for the short and long CEN
exposures, as well as for the HALF and parallel (background) fields
taken close in time.  For the STIS fields, we used a single AC for each
of our cluster fields, which we found to be very similar to those
obtained by Houdashelt, Wyse \& Gilmore (2001).  We list our aperture
corrections in Table \ref{apcor.tab}. 

\input{apcor.tex}

Before applying these corrections, however, we first corrected the
aperture magnitudes for the geometric distortion of the {\sl WFPC2}
chips, using the correction equations from Holtzman et al.  (1995a), and
then for the time-dependent charge transfer (in)efficiency (CTE)
determined by Whitmore, Heyer \& Casertano (1999). 

\subsection{Photometric Calibration}
\label{calib.sect}

\subsubsection{WFPC2 photometry}

We used the transformation coefficients of Holtzman et al.\ (1995a) to
convert the aperture-corrected {\sl WFPC2} photometry to the standard
Johnson-Cousins {\it V, I} photometric system:
\begin{eqnarray}
\label{standV.eq}
V &=& -2.5 \times \log \dot{C}({\rm F555W}) +
(-0.052 \pm 0.007) \times (V-I) \nonumber \\
&+& (0.027 \pm 0.002) \times (V-I)^2 + (21.725 \pm 0.005) \nonumber \\
&+& 2.5 \times \log ({\rm GR}) \, , 
\end{eqnarray}   
and
\begin{eqnarray}
\label{standI.eq}
I &=& -2.5 \times \log \dot{C}({\rm F814W}) +
(-0.062 \pm 0.009) \times (V-I) \nonumber \\
&+& (0.025 \pm 0.002) \times (V-I)^2 + (20.839 \pm 0.006) \nonumber \\ 
&+& 2.5 \times \log ({\rm GR}) \, .
\end{eqnarray}
Here, $\dot{C}$ is the count rate in $0.''5$ apertures, and GR
is the gain ratio as defined by Holtzman et al.  (1995b), which accounts
for the difference in gain states between calibration and science
observations.  For {\sl WFPC2} in its state with an analog-to-digital
gain of 7 electrons (bay 4; as for our observations) GR is 1.987, 2.003,
2.006 and 1.955 for the PC, WF2, WF3 and WF4 chips, respectively,
subject to an $\sim 1$\% uncertainty.  These transformations hold for
$-0.3 < (V-I) < 1.5$, which covers most of the colour range of the stars
in our LMC clusters.  The colour terms in Eqs.\ (\ref{standV.eq}) and
(\ref{standI.eq}) are defined in the standard system; consequently, the
equations must be applied iteratively to measures from the {\sl WFPC2}
frames. 

We found a small offset between source magnitudes from the long and
short CEN exposures (cf.  Johnson et al.  2001).  Since the main purpose
of the short CEN exposures was to obtain reliable magnitudes for the
brightest stars populating the cluster LFs, we applied a simple
magnitude offset to the short exposures.  These offsets, which we have
listed in Table \ref{offsets.tab}, were determined over the common range
of magnitudes in the short and long exposures with similar measurement
uncertainties (based on the appearance of their CMDs), i.e.  in the
range $18.6 \le {\rm mag} \le 20.0$ and $18.4 \le {\rm mag} \le 19.8$
for NGC 1805 and NGC 1818, respectively, for both {\it V} and {\it I}. 
Since they were determined independently, and using a different method
of analysis, small differences between our and Johnson et al.'s (2001)
offsets occur, although not in a systematic way.  This lends additional
credibility to our photometry. 

\input{offsets.tex}

At this point, we combined the master source lists obtained from the CEN
long and short exposures, and subsequentially the lists from the
combined CEN exposures and the HALF fields. The merging of the short and
long CEN exposures was based on the following considerations:

\begin{enumerate}

\item If both long and short magnitudes are available for a particular
object, use the long exposures, but only if $V_{\rm long} > V_{\rm
sat.,long}$ {\it and} $I_{\rm long} > I_{\rm sat.,long}$, where the
subscript ``sat.,long'' indicates the saturation limit of the long
exposures (Table \ref{sources.tab}).  Otherwise, use the short
exposures, but only if the observations are not saturated in at least
one passband. 

\item If only magnitudes from the long exposures are available for the
object, use these if the observations are not saturated in at least one
passband. 

\item If only short-exposure magnitudes are available, use these if
either $V_{\rm short} < V_{\rm sat.,long}$ or $I_{\rm short} < I_{\rm
sat.,long}$, {\it and} the observations are not saturated in at least
one passband. 

\end{enumerate}

Before combining the CEN and HALF field detections, we also needed to
correct the HALF exposures for the small photometric offset found for
matching sources in the CEN fields; the corresponding offsets are also
listed in Table \ref{offsets.tab}.  The subsequent merging of the CEN
and HALF fields was done largely along similar lines:

\begin{enumerate}

\item If only magnitudes from the combined CEN list are available, keep
these.

\item If both CEN and HALF magnitudes were determined, keep the CEN
magnitude if it is brighter than $V = 23.0$ and $I = 23.5$; otherwise
use the HALF data.

\item If only a HALF-field detection was registered, keep this if the
object is not saturated in at least one passband.

Finally, we corrected our photometry for the effects of Galactic
foreground extinction using the extinction values tabulated in Table
\ref{sample.tab}.  Most of these values were derived from isochrone fits
to the clusters' CMDs (Castro et al.  2001, SBJG).  We determined the
extinction in the {\sl HST} passbands assuming the Galactic extinction
law of Rieke \& Lebofsky (1985), convolved with the {\sl HST} filter
shapes, $A_{\rm F555W} / A_V = 1.081$ and $A_{\rm F814W} / A_V = 0.480$
(de Grijs et al.  2001).  One should be cautious to apply an extinction
estimate for a non-standard bandpass, which is then subsequently
transformed to a standard magnitude.  For NGC 1805, however, the
difference in the extinction correction between the F555W and {\it
V}-band filters is negligible for practical purposes, namely $\le 0.010$
mag, while for the F814W/{\it I} band this reduces to $A_I - A_{\rm
F814W} \le 0.001$ mag.  The equivalent extinction differences for NGC
1818 are $\le 0.007$ and $\le 0.001$ mag, respectively. 

\end{enumerate}

\subsubsection{STIS photometry}

To convert our STIS LP magnitudes to the standard {\it V} magnitudes, we
first transformed them to {\sl WFPC2} flight system magnitudes using the
empirical transformations obtained by Beaulieu et al.  (2001) for STIS
CCD observations with analog-to-digital gain 4 (as for our
observations):
\begin{eqnarray}
\label{stisV.eq}
V_{555} &=& 23.473 - 2.5 \times \log \dot{C} - 0.5184 \times (V_{555} -
I_{814}) \nonumber \\
&-& 0.0502 \times (V_{555} - I_{814})^2 \, ,
\end{eqnarray}
and subsequently used the standard {\sl WFPC2}-to-ground transformations
to obtain standard {\it V} magnitudes (cf.  Holtzman et al.  1995a). 
The STIS-to-$V_{555}$ transformation was derived for the Galactic GC
NGC 6553 ([Fe/H] $\approx -0.2$) for magnitudes determined in $0.5''$
apertures, and are valid for sources with $1.8 \lesssim (V_{555} -
I_{814}) \lesssim 4$. 

Houdashelt et al.  (2001) also obtained empirical transformations
between STIS LP and {\sl WFPC2} flight magnitudes, based on the Galactic
GCs 47 Tuc ([Fe/H] $\sim -0.7$) and M15 ([Fe/H] $\sim -2.2$), using the
F606W and F814W {\sl WFPC2} filters.  They concluded that -- if
quadratic transformations are assumed (as in Beaulieu et al.  2001) --
there is no evidence for a metallicity dependence in the transformation
relations.  Their transformation equations apply to the entire $(V_{606}
- I_{814})$ colour range covered by their observations, i.e.  $0.35
\lesssim (V_{606} - I_{814}) \lesssim 1.85$, which translates to $-0.1
\lesssim (V_{555} - I_{814}) \lesssim 1.4$ for stellar spectral types in
the range from early B to G, which is based on the folding of synthetic
spectra through the filter bandpasses using the {\sc synphot} package in
IRAF. 

Since a significant number of the STIS-detected sources did not have
counterparts with well-determined magnitudes in either the F555W or
F814W images, we assumed all sources in the final STIS source list to be
cluster stars and approximated their $(V_{555} - I_{814})$ colours to be
those of the main sequence stars of the appropriate magnitude determined
from our {\sl WFPC2} CMDs. 

To correct for foreground extinction, we used the extinction correction
obtained by Beaulieu et al.  (2001), $A_{V_{\rm LP}} = 2.505 E(B-V)$,
which is based on integration across the passband. 

\subsection{Completeness}
\label{compl.sect}

Due to the significant stellar density gradient across the cluster
fields, completeness corrections are a strong function of position
within a cluster.  Therefore, we computed completeness corrections for
all observations in circular annuli around the centre of each cluster,
for both the PC and the WF fields, located at intervals between the
centre and $3.6''$, $3.6-7.2''$, $7.2-18.0''$, $18.0-36.0''$,
$36.0-54.0''$ and at radii $\ge 54.0''$.  At the distance of the LMC,
$1''$ corresponds to $\sim 0.25$ pc.  The cluster centre coordinates,
included in Table \ref{sample.tab}, were determined by smoothing the
cluster light distributions and subsequently applying an ellipse-fitting
routine to the smoothed cluster profiles. 

\begin{figure*}
\psfig{figure=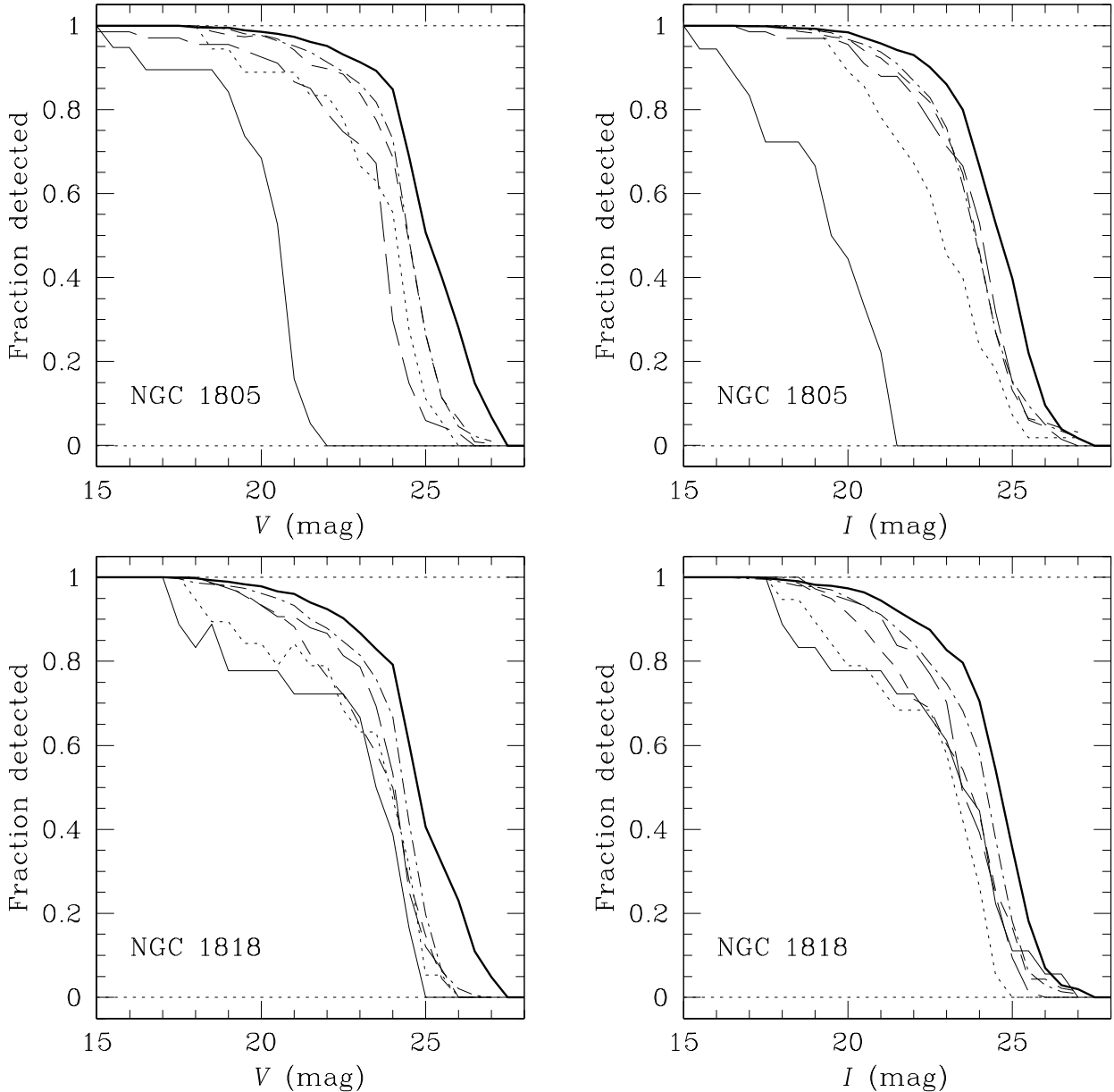,width=18cm}
\caption{\label{compl.fig}Completeness curves for NGC 1805 and NGC 1818. 
The different line styles refer to different annuli: thin solid -- $r
\le 3.6''$ (NGC 1805: short exposures); dotted -- $3.6'' < r \le 7.2''$;
long dashes -- $7.2'' < r \le 18''$; short dashes -- $18'' < r \le
36''$; dash-dotted -- $36'' < r \le 54''$; thick solid -- $r > 54''$.}
\end{figure*}

We added an area-dependent number of Gaussian sources to the individual
annuli, ranging from $\sim$ 60 in the inner annulus to 2000 in the
outer, partial annulus.  We created artificial source fields for input
magnitudes between 15.0 and 27.0 mag, in steps of 0.5 mag.  Their
$(V-I)$ colours were distributed following the clusters' main sequence
ridge lines, i.e., they were magnitude dependent.  We then applied the
same source detection routines to the fields containing the combined
cluster stars and the artificial sources.  The results of this exercise,
based on the long CEN exposures, are shown in Fig.  \ref{compl.fig}. 
These completeness curves were corrected for the effects of blending or
superposition of multiple randomly placed artificial stars as well as
for the superposition of artificial stars on genuine objects.  Due to
the large number of bright, saturated stars in the innermost annulus of
NGC 1805, proper completeness tests could not be done in this area; for
comparison, we have plotted the corresponding completeness curve in this
area for the short CEN exposures instead.  The progressive increase in
completeness fraction with radius for a given source brightness clearly
illustrates the potentially serious effects of crowding in the inner
regions of the clusters.  In the analysis performed in this paper, we
only include those ranges of the stellar LF where the completeness
fraction is in excess of 50\%. 

As mentioned in Sec.  \ref{sources.sec}, the {\sl HST} PSF is
non-Gaussian in shape.  Therefore, we have introduced an additional
uncertainty by adding Gaussian-shaped artificial objects to our science
frames for the completeness analysis.  A detailed comparison of the
light profiles of the Gaussian sources on the one hand and PSFs from the
{\sl WFPC2} PSF library and Tiny Tim artificial PSFs on the other,
reveals that, except for the innermost pixel, both light profiles
closely match each other.  Since our source detection routine requires
information on both the brightness {\it and} the shape of the objects in
order to include them, and because for those magnitudes where the
incompleteness becomes significant source detections are photon noise
limited and therefore relatively independent of the stellar profile
shape, we estimate that by using Gaussian light profiles, we will have
obtained {\it conservative} completeness estimates.  In other words, for
any given magnitude range, we may have obtained completeness fractions
that are marginally too low, and would need to be shifted to slightly
fainter magnitudes ($\ll 0.5$ mag). 

\begin{figure*}
\psfig{figure=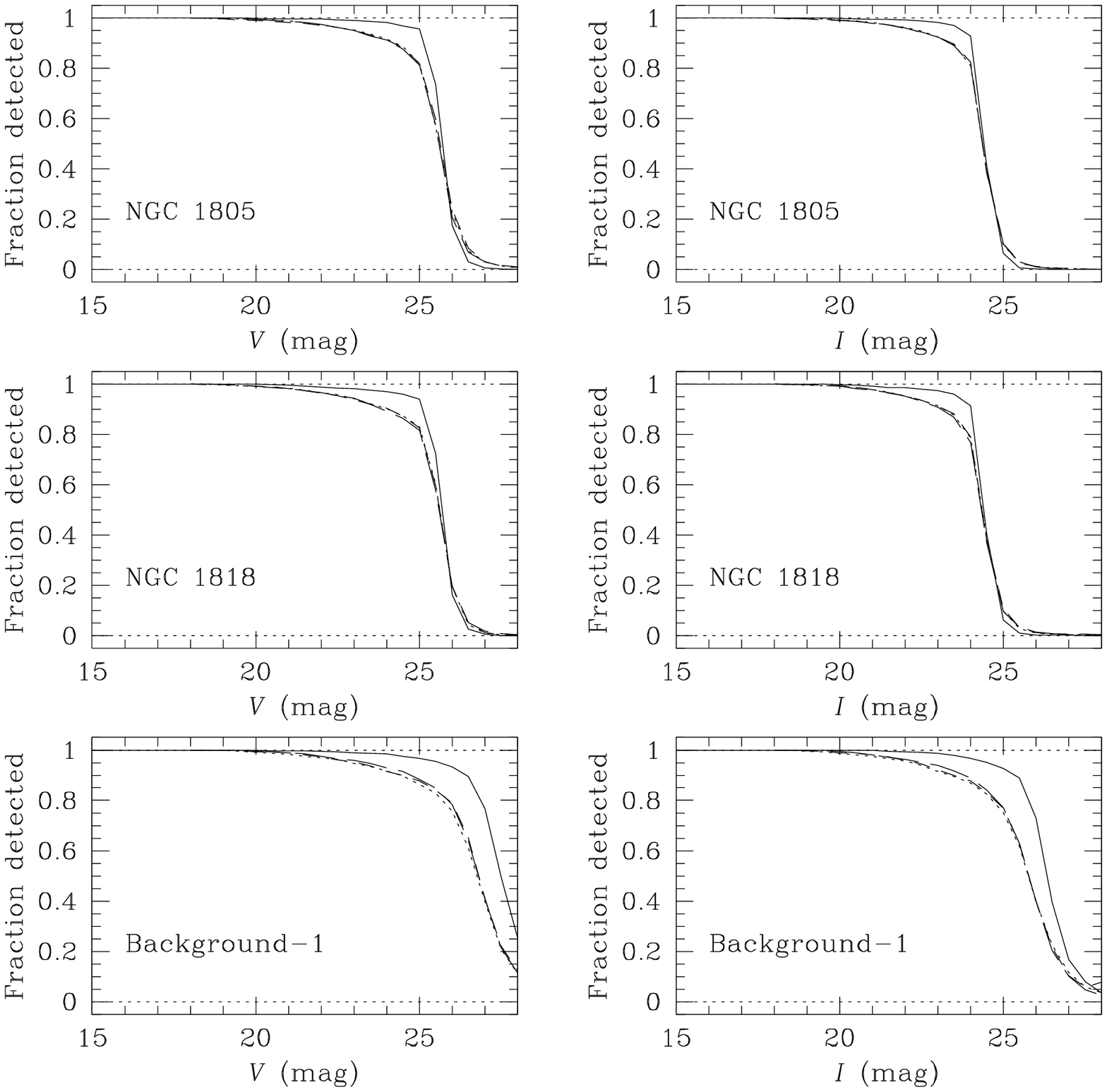,width=18cm}
\caption{\label{complbg.fig}Completeness curves for the background
fields.  The different line styles refer to the individual {\sl WFPC2}
chips; the smaller pixel size of the PC chip (solid lines) clearly
increases the completeness fraction; for the WF chips (dotted, long and
short-dashed lines), a single completeness correction applies.}
\end{figure*}

We also determined completeness curves for the background fields. 
Thanks to their low stellar density, a single completeness curve for the
WF chips was found to apply to any one background field, as shown in
Fig.  \ref{complbg.fig}.  The higher resolution obtained with the PC
translates into a greater completeness fraction compared to the WF
chips.  The completeness curves for the NGC 1805-par and NGC 1818-par
fields are identical, within the observational uncertainties. 

Finally, we applied the same analysis to the STIS science frames and
background fields, using a single completeness curve for the entire STIS
field of view.  The 50\% completeness limit occurs at $V_{555} \simeq
25$ and $\simeq 24.5$ in NGC 1805 and NGC 1818, respectively, for the
science frames and -- due to their shorter exposure times -- roughly
0.2--0.3 mag brighter in the parallel background fields. 

\section{Approach: data preparation and luminosity functions}
\label{approach.sec}

In this section, we will examine the dependence of the shape and slope
of the stellar LF on position within the clusters.  We basically use the
cluster LFs as smoothing functions of the full two-dimensional ({\it V,
V--I}) CMDs.  The full {\it N}-body modeling of the cluster CMDs, fully
inclusive of the latest input physics (e.g., binary star mergers), {\it
except the effects of mass segregation}, was discussed by Johnson et al. 
(2001) and shown to represent the observed CMDs very well.  Where the
full 2-dimensional CMDs were used to infer the presence and the effects
of mass segregation, this was mostly based on differences in the
concentration of specific stellar types, most often main sequence and
giant branch stars, in young LMC clusters (cf.  Sect. 
\ref{lmcclus.sect}) and old GCs (e.g., in NGC 1851, Saviane et al. 
1998; NGC 5466, Nemec \& Harris 1987; Pal 12, Harris \& Canterna 1980;
and 47 Tuc, Da Costa 1982).  However, our cluster stars start to
saturate at the faint end of the red giant branch, so this approach is
not feasible.  In fact, with the exception of a handful of the brightest
stars, we are limited to the analysis of the main sequence stars in
these clusters (cf.  the CMDs in Johnson et al.  2001). 

In order to study the positional dependence of the LF within our star
clusters, we need to correct the observed stellar LFs in the CEN and
HALF fields for the contribution of stars from the galactic (LMC)
background in these fields.  For that purpose, we obtained the
background field LFs using identical procedures as for the CEN and HALF
fields themselves.  The resulting LFs are statistically
indistinguishable from Castro et al.'s (2001) LFs, although small
non-systematic differences occur due to the different methods used.  We
combined the LFs obtained from the three individual fields into a single
background LF.  The uncertainties due to possible background variations
thus introduced are negligible compared to the errors in our source
magnitudes, in particular for the fainter stars, where such background
variations would be of importance (cf.  Castro et al.  2001). 

The LFs needed to be scaled to apply to the appropriate areas used to
study the radial dependence of the cluster LFs, and subsequently
subtracted from those.  Since all of our background fields have longer
exposure times than the longest CEN exposures, saturation sets in at
brighter magnitudes than for these CEN fields.  To correct for the
background contribution at the brightest magnitudes, $V \le V_{\rm
sat.,bg}$, and $I \le I_{\rm sat.,bg}$, we simply extrapolated the
background field LFs to $V = V_{\rm sat.,short}$, and $I = I_{\rm
sat.,short}$, i.e., to the brightest unsaturated stars in our short CEN
exposures.  The extrapolated number of background field stars brighter
than the saturation limits of our background fields is negligible for
practical purposes, however.  In the remainder of this paper, when we
refer to the cluster LFs, this applies to the background-corrected LFs. 

Foreground stars are not a source of confusion in the case of our LMC
clusters.  From careful inspection of the CMDs presented by Johnson et
al.  (2001), we conclude that for $V \lesssim 23$ there are most likely
no foreground stars in our fields of view.  This is consistent with the
standard Milky Way star count models (e.g., Ratnatunga \& Bahcall 1985). 

\subsection{Evidence for systematic luminosity segregation}
\label{lumsegr.sect}

\begin{figure*}
\psfig{figure=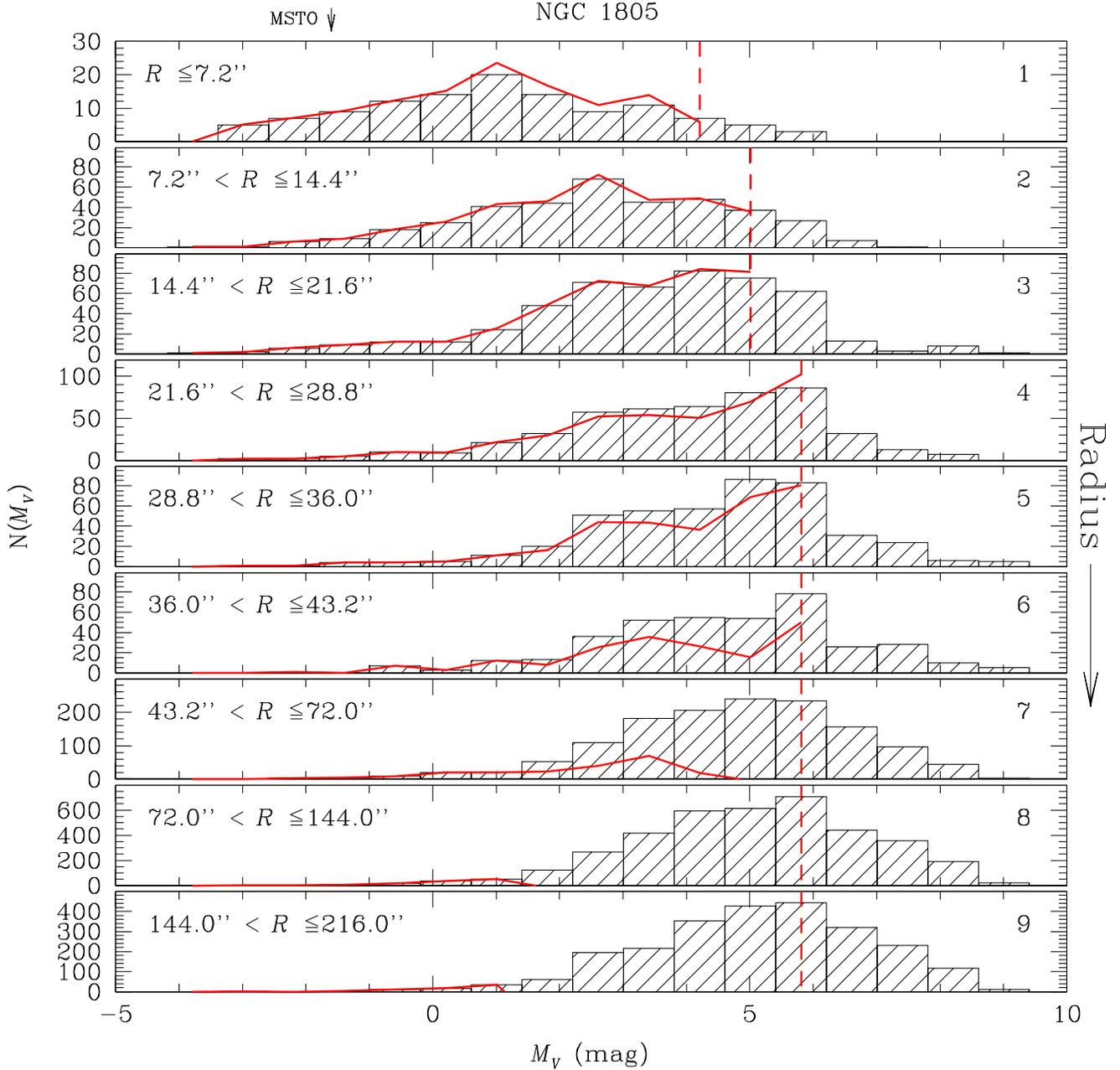,width=18cm}
\caption{\label{n1805hist.fig}Observational total LFs from our {\sl
WFPC2} data in annuli at increasingly large radii from the NGC 1805
cluster centre (histograms).  The thick solid lines are the actual
cluster star distributions, after correction for the background field
star contribution and the effects of incompleteness; the 50\%
completeness limits are indicated by the vertical dashed lines. The
numbers on the right-hand side in each panel refer to the corresponding
annular LFs in Fig. \ref{lfslopes.fig}. The main sequence turn-off
magnitudes (``MSTO'') are indicated by the arrows.}
\end{figure*}

\begin{figure*}
\psfig{figure=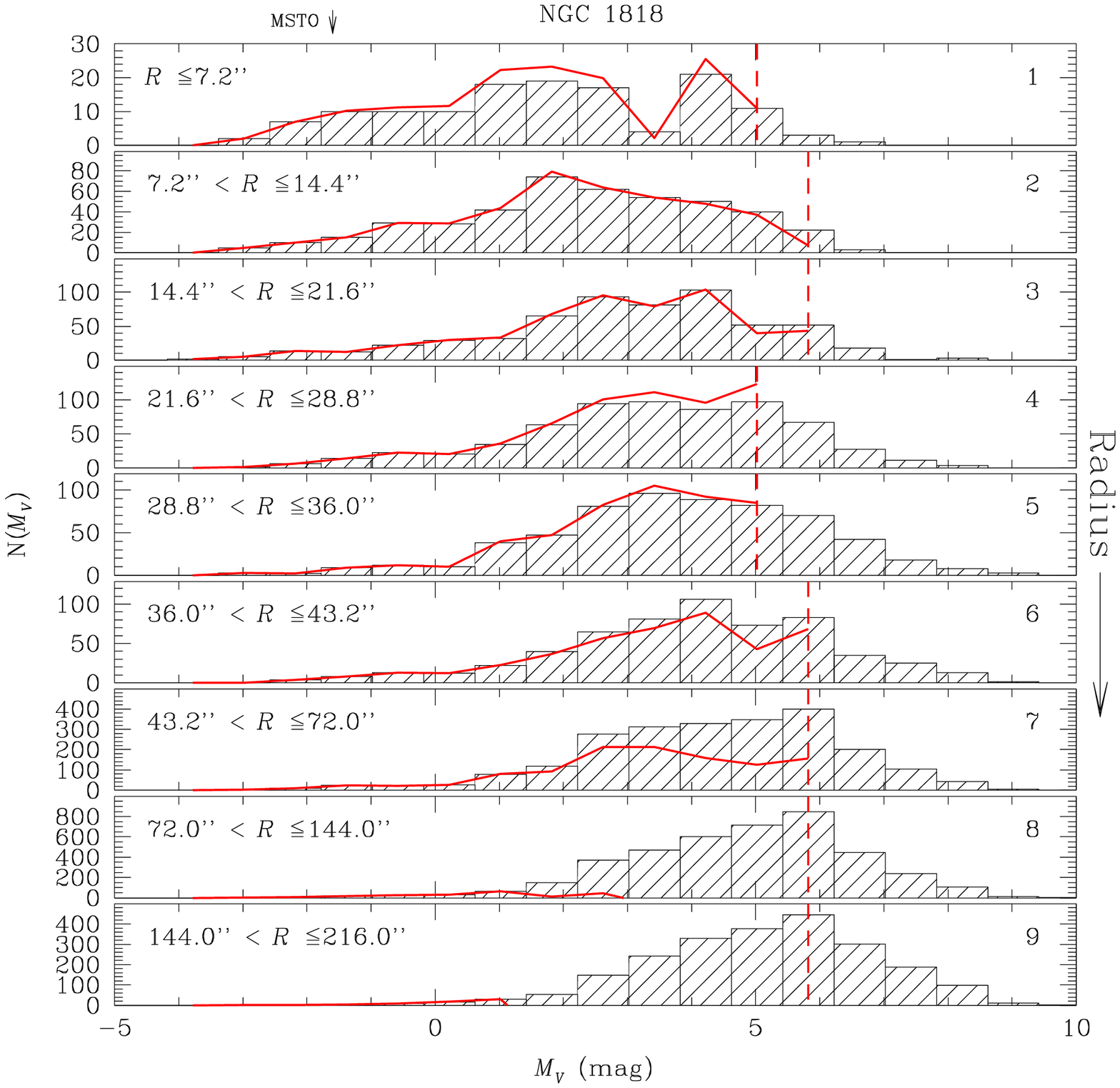,width=18cm}
\caption{\label{n1818hist.fig}Same as Fig. \ref{n1805hist.fig}, but for
NGC 1818.}
\end{figure*}

Figs.  \ref{n1805hist.fig} and \ref{n1818hist.fig} show the distribution
of stellar magnitudes as a function of distance from the cluster
centres.  The shaded histograms represent the total number of stars in
our final source lists, not corrected for incompleteness, area covered
or background star contamination; the thick solid lines are the actual
cluster star distributions, obtained by subtracting the background
contribution expected in the area covered by each annulus from the
observed total LFs and subsequently correcting for incompleteness
effects.  The 50\% completeness limits in each annulus are indicated by
the vertical dashed lines through the centres of the last magnitude bin
above this limit.  We have used the {\it V}-band completeness curves to
correct our LFs for the effects of incompleteness.  However, since the
final source lists of cluster stars were based on cross referencing the
detections both in the {\it V} and {\it I} bands, we need to be careful
close to the 50\% completeness limit in {\it V}.  Close examination of
Fig.  \ref{compl.fig} shows that the {\it I}-band magnitude is always
brighter than the corresponding {\it V}-band magnitude, for any
completeness fraction.  This affects the derived LFs in Fig. 
\ref{n1805hist.fig}, in particular for $M_V \lesssim 2.7$ ($R \le
7.2''$) and $M_V \lesssim 3.5$ ($7.2 < R \le 14.4''$), so that any LF
slope derived for these radial ranges and including stars within $\sim
1.5$ magnitude of the 50\% completeness limit is a {\it lower} limit. 
This effect does not play a significant role for NGC 1818. 

Secondly, for the inner radial bin ($R \le 7.2''$) of NGC 1805 the
incompleteness for the fainter stars is severe due to the large
concentration of bright stars in the very compact core of this cluster. 
Therefore, the declining LF for $M_V \gtrsim 1$ is likely due to the
central incompleteness in this inner radial bin, and any LF slope
derived for these stars will thus also be a lower limit. 

The effects of strong luminosity segregation are clearly visible, in the
sense that the brighter cluster stars are strongly concentrated in the
inner $\sim 20''$ in each cluster.  The brightest unsaturated cluster
stars, i.e., $V \lesssim 17$, consist of a mixture of stars at the top
of the main sequence and giant branch/red clump stars (cf.  the CMDs in
Johnson et al.  2001), which span only a very narrow mass range, while
the fainter stars are main sequence stars.  The effects are strongest in
NGC 1805; strong luminosity segregation in this cluster was already
apparent from an initial visual examination of Fig.  \ref{images.fig}. 

\begin{figure*}
\psfig{figure=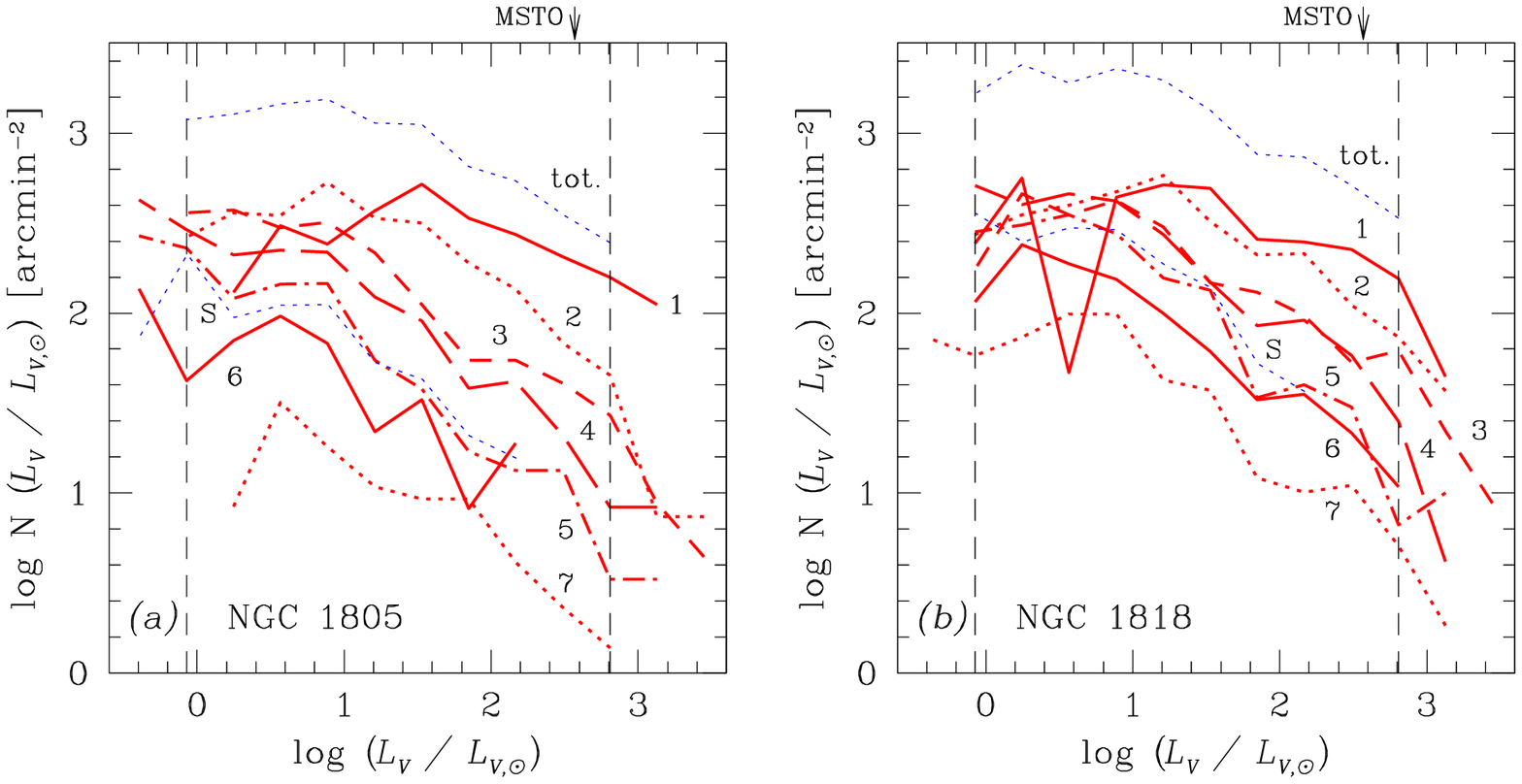,width=18cm}
\vspace*{-9cm}
\caption{\label{lfslopes.fig}Luminosity functions in NGC 1805 and NGC
1818: comparison of annular LFs from the inner 7 annuli shown in Figs. 
\ref{n1805hist.fig} and \ref{n1818hist.fig} and the STIS HALF fields,
normalised to 1 arcmin$^2$ area coverage.  For reasons of clarity, we
have omitted the vertical error bars.  The dotted lines covering the
range between the fitting boundaries (vertical dashed lines) and
representing the largest numbers of stars / arcmin$^2$ are the overall
cluster LFs.  The annular bins are numbered as in Figs. 
\ref{n1805hist.fig} and \ref{n1818hist.fig}, and represented by the
following line styles (inside outwards), from top to bottom: solid,
dotted, short dashed, long dashed, dot-dashed, dotted (STIS HALF fields,
``S''), solid, and dotted.  The approximate main sequence turn-off
(``MSTO'') luminosities are indicated by the arrows.}
\end{figure*}

In Figs.  \ref{lfslopes.fig}a and b we show all annular cluster LFs out
to $R = 72.0''$, corrected for the effects of incompleteness (as a
function of radial distance from the cluster centres), background
contamination and for the sampling area covered by each (partial)
annulus, for NGC 1805 and NGC 1818, respectively.  In this
representation, the radial dependence of the cluster LFs is more easily
visible than in Figs.  \ref{n1805hist.fig} and \ref{n1818hist.fig}.  For
the NGC 1818 field, these annular LFs are not contaminated by the
companion cluster NGC 1818 B, since the latter is located at larger
radii than covered by our LFs.  We also show the overall cluster LFs,
constructed by adding all individual annular LFs weighted by the area
covered by the observations.  This procedure ensures the proper
treatment of the sampling incompleteness at each annular radius.  For
NGC 1805 the completeness corrections of the innermost annulus (and to
some lesser degree also of the second annulus) are uncertain due to
severe crowding and statistical noise (small-number statistics), likely
causes an artificial turndown for the lowest luminosities.  However, the
contribution of this annular LF to the overall LF is $\lesssim 5$\%,
because of the weighting by area used in its construction, with the
inner annulus covering only $\sim 1.7$\% of the total area covered by
our observations for $R \le 72.0''$

In paper III we will extend the overall LFs down to the
lowest-luminosity sources detected reliably in our STIS fields, for the
entire LMC cluster sample.  Our ultimate aim is to determine whether the
cluster LFs -- and thus their MFs -- are statistically indistinguishable
or significantly different over the entire mass range, down to the
lowest masses.  This forms part of our efforts to determine the strength
of the apparent universality of the IMF. 

We subsequently determined the LF slopes, assuming a simple power-law
dependence, i.e., $N(L) \propto L^{-\alpha}$, where $\alpha$ is the LF
slope (but see Sec.  \ref{discussion.sect}).  Although we used the
power-law approximation to be able to compare our results to previously
published LF slopes, we realise that the inner LFs in Figs. 
\ref{n1805hist.fig} and \ref{n1818hist.fig} show a clear maximum inside
our fitting ranges, and that the overall cluster LFs are clearly {\it
not} linear.  Despite this, a comparison of LF slopes obtained using
power-law fits over identical luminosity ranges is still valuable to
quantify the radial dependence of the cluster LFs, however.  We chose to
use fitting ranges in luminosity that covered the maximum overlap among
our annular LFs in order to minimise the effects of small-scale
statistical fluctuations in the LFs.  For NGC 1805, we used the ranges
$-2.2 \le M_V \le 4.2$ ($2.81 \ge \log (L_V / L_{V,\odot}) \ge 0.25$)
for $R \le 72.0''$ and the slightly greater common magnitude range $-2.2
\le M_V \le 5.0$ ($2.81 \ge \log (L_V / L_{V,\odot}) \ge -0.07$) for
$7.2 < R \le 72.0''$.  The latter magnitude range was also used for
fitting the LF slopes in NGC 1818, and is indicated by the vertical
dashed fitting boundaries in Fig.  \ref{lfslopes.fig}.  We note,
however, that Johnson et al.'s (2001) CMDs show confusion by Be stars
for $V \lesssim 17, M_V \lesssim -1.6$, so that our brightest data point
needs to be taken with caution.  This does not affect the overall
results presented in this section, however.  The error bars in the LF
slopes include the formal error and the statistical uncertainties due to
poisson noise. 

\begin{figure*}
\psfig{figure=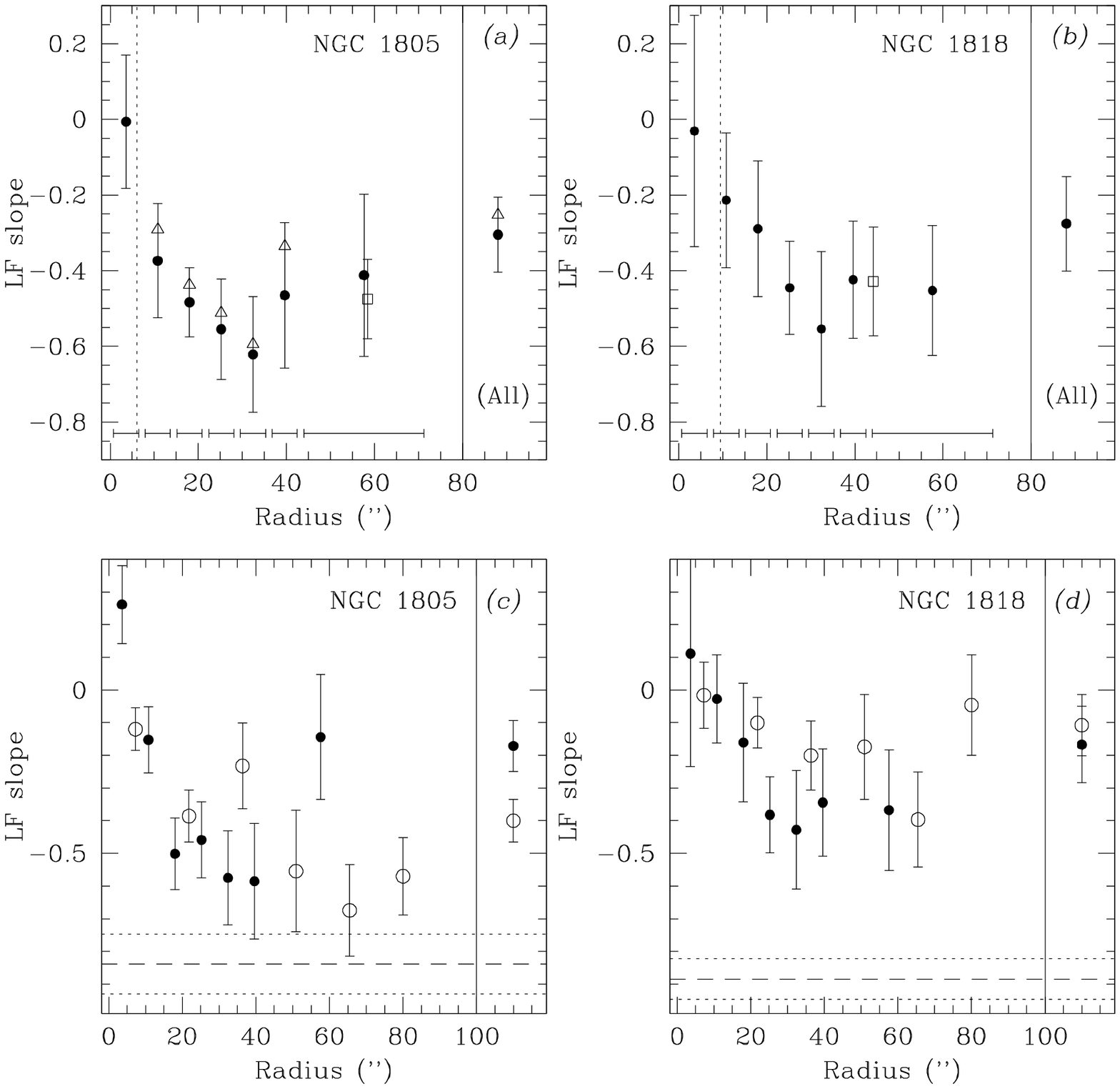,width=18cm}
\caption{\label{litcomp.fig} The dependence of the LF slope on the
fitting range.  {\it (a)} and {\it (b)} LF slopes determined between the
fitting boundaries indicated in Fig.  \ref{lfslopes.fig}.  (NGC 1805:
the filled circles and error bars represent the smaller luminosity
range, which could be applied to the entire radial cluster distribution;
the open triangles were obtained by using the greater range in
luminosity).  We have also indicated the slope one would measure based
on the overall cluster LFs (right-hand subpanels), and the slope
obtained from the STIS HALF field CLFs (open squares).  The vertical
dotted lines indicate the cluster core radii.  Comparison of our LF
slopes with published in the literature.  {\it (c)} and {\it (d)} --
Comparison with SBJG after redetermination of the LF slopes using
identical absolute magnitude ranges for each sample.  Black dots: this
paper, open circles: SBJG.  The horizontal dashed and dotted lines
represent the sky background level and its 1-$\sigma$ uncertainty,
respectively, as discussed in the text.}
\end{figure*}

In the top panels of Fig.  \ref{litcomp.fig} we show the dependence of
the slope of the LF on cluster radius; the radial ranges to which the
data points apply are indicated by small bars at the bottom of each
panel. 

We have also included the LF slopes from the STIS HALF field LFs (open
squares), obtained following identical procedures as for the {\sl WFPC2}
{\it V}-band LFs.  We used a fitting range to obtain the LF slope of
$-1.0 \le M_V \le 5.0$ ($2.33 \ge \log (L_V / L_{V,\odot}) \ge -0.07$)
for both of our single fields.  The source distributions in our STIS
fields peak at 58.4$''$ and 44.2$''$ for NGC 1805 and NGC 1818,
respectively.  Their radial extent is roughly 48 and 44$''$,
respectively (i.e., the FWHMs of the radial distributions).  The fact
that the STIS data points are entirely consistent with the {\sl WFPC2}
results, although they were determined completely independently and
using a different detector, confirms the robustness of these results. 
For comparison, we have also indicated the LF slopes one would arrive at
by taking the overall cluster LFs, indicated by ``all''. 

Both clusters show clear evidence of luminosity segregation, in the
sense that the LF slopes steepen with increasing cluster radius.  In NGC
1805, the LF slopes reach a stable level beyond $\simeq 15''$ (or $\sim
3.8$ pc), well beyond the cluster's half-light radius at 1.8 pc.  A
stable LF slope is reached for $R \gtrsim 25''$ ($\sim 6.3$ pc) in NGC
1818, again indicating strong luminosity segregation in the inner
annuli. 

Although the trend towards steeper LFs with increasing radius is clear,
the associated error bars are large.  They are the formal errors from
the fit of a single power law to the data points, and are therefore
dominated by the non-linearity of the annular LFs and point-to-point
variations.  The non-linear behaviour of the annular LFs is also our
preferred explanation for the $\sim 10$\% difference in measured LF
slopes for our two fitting ranges in luminosity used for NGC 1805.  We
will return to this issue in Sec.  \ref{discussion.sect}. 

\subsection{The brightest cluster stars}

\begin{figure}
\psfig{figure=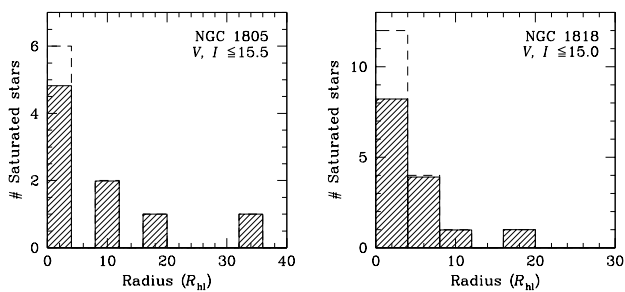,width=9cm}
\vspace{-4.5cm}
\caption{\label{saturated.fig}Radial distributions of the brightest
(saturated) stars in NGC 1805 and NGC 1818; dashed histogram: original
star counts, shaded histograms: corrected for the clusters' surface
brightness distribution, as described in the text.}
\end{figure}

Finally, we can use the distribution of the saturated stars in the short
CEN exposures to strengthen our conclusions on the presence of strong
luminosity segregation in both clusters.  Saturated stars in the short
CEN exposures are brighter than 15.5, and 15.0 mag (in both {\it V} and
{\it I}) in NGC 1805 and NGC 1818, respectively.  Fig. 
\ref{saturated.fig} shows the radial distribution of these brightest
cluster members, in units of the clusters' half-light radii. We show the
original star counts as the dashed histograms. However, to interpret
these distributions in terms of mass segregation, we need to correct
these for the underlying cluster surface brightness distribution. We
assumed King-like cluster profiles of the form suggested by Elson et al.
(1987) to obtain the corrected, shaded histograms.

We can firmly rule out a Galactic foreground origin for these sources,
since the standard Milky Way models predict $\lesssim 0.005$ foreground
stars to appear in an area corresponding to the PC field of view towards
these clusters.  There is clear evidence for luminosity segregation in
both clusters, with a strong concentration of the brightest stars within
the inner $\sim 4$ and $\sim 8 R_{\rm hl}$ in NGC 1805 and NGC 1818,
respectively.  This result is largely independent of the radial bin size
adopted. 

\subsection{Ejection of bright stars?}
\label{nttfields.sect}

Following similar procedures as for the {\sl WFPC2} observations, we
obtained calibrated {\it V} and {\it I}-band source lists for the
wide-field NTT data.  These observations were obtained to study the
distribution of cluster (and field) stars near and above the main
sequence turn-off, which occurs at $V \sim 17$ for both clusters.  The
magnitude range $15.5 \lesssim V \lesssim 20.0$ is available for the
combined cluster and field star populations. 

The main purpose of our use of the NTT fields was to establish whether
there is evidence that a fraction of the brighter cluster stars have
been ejected from the cluster on time-scales similar to the cluster
ages, $\sim 10 - 25$ Myr.  Therefore, we examined the radial
distributions of the combined cluster+field star populations (centred on
the cluster centres) for 0.5 mag bins, ranging from $V = 15.5$ to $V =
20.0$.  Based on Figs.  \ref{n1805hist.fig} and \ref{n1818hist.fig}, we
conclude that for radii $R \gtrsim 72.0''$ the LFs are dominated by the
background field population.  We will therefore conservatively focus our
analysis of the NTT fields on radial 100 pixel ($27''$) bins beyond $R =
100''$, out to the edge of our fields of view at $\simeq 330''$.  For
stars fainter than $V \sim 17.5$, the numbers of stars as a function of
radius, corrected for the area covered, are consistent with a flat
distribution.  This indicates that for these magnitude bins the observed
stellar population is largely dominated by the background field. 
Therefore, we decided to use the radial distribution of stars in the
magnitude range $18.5 \le V \le 19.5$ ($0 \lesssim M_V \lesssim 1$) as
our control sample, to be used for the normalisation of the brightest
magnitude ranges.  At fainter magnitudes, one needs to take significant
incompleteness corrections into account. 

Although the numbers of bright stars in each of our fields are small (up
to several tens in each 0.5 mag bin, for the entire radial range from
$100'' - 330''$), for NGC 1818 there appears to be an excess of $\sim
8-10$\% of brighter stars (i.e., an excess of $18 \pm 8$ stars of $15.5
\le V \le 16.0$, as well as brighter, saturated stars) towards the inner
radial cut-off at 100$''$ compared to the control sample.  Because of
the relatively small numbers of bright stars, and the associated
statistical uncertainties, this is a $\sim 2 \sigma$ result.  This
overdensity is consistent with a radially decreasing distribution,
suggesting that these bright stars are associated with NGC 1818.  We
will discuss the implications of this in terms of formation scenarios or
dynamical ejection from the cluster core in Sec.  \ref{ejections.sect}. 

For stars with magnitudes $V > 16.5$ in NGC 1818 and for the entire
range in magnitudes for NGC 1805, the radial distributions are
consistent with randomly distributed stars in the field, again compared
to our control sample, $18.5 \le V \le 19.5$. 

Finally, for the NTT field of view used for this analysis ($540'' \times
509''$), the Galaxy models of Ratnatunga \& Bahcall (1985) predict
roughly 14 foreground stars in the range $15 \le V \le 17$.  Our samples
contain at least twice as many stars in this magnitude range, for either
field. In addition, if a significant fraction of the stars in this
magnitude range were Galactic foreground stars, they would not be
expected to be concentrated on the clusters, as found for NGC 1818.

\section{Discussion}
\label{discussion.sect}

\subsection{Comparison with previously published results}

In this section, we compare our results to those of SBJG to illustrate
the sensitivity of a simple single-parameter fit of an LF to star count
data for compact star clusters.  All of the adopted luminosity range,
radial range, completeness range, and background subtraction affect an
apparently robust result. 

Since SBJG's published annular LF slopes were determined over a
different magnitude fitting range than ours, we redetermined the slopes
for both our LFs and those of SBJG over the common magnitude range $0.25
\le M_V \le 4.25$ ($1.83 \ge \log (L_V / L_{V,\odot}) \ge 0.23$), after
converting SBJG's {\sl WFPC2} flight system magnitudes to the standard
{\it V}-band system.  The results are shown in Figs.  \ref{litcomp.fig}c
and d, where the black dots represent the slopes determined from the LFs
derived in this paper and the open circles are those from SBJG's
published figures.  The right-hand subpanels show the LF slopes if we
had considered the entire stellar distribution of the clusters at once. 
We observe reasonable consistency between our results, within the
associated fitting uncertainties, although a small discrepancy is seen
in the outer regions, beyond $R \simeq 50''$ for NGC 1805 and $R \simeq
70''$ for NGC 1818.  Small differences between the overall cluster LFs
of NGC 1805 could be explained by the fact that we used a radial range
$R \le 72.0''$, while SBJG's overall LF is based on the cluster members
out to $R = 102.0''$. 

Although the consistency between both our results is encouraging, in
particular since they were obtained entirely independently, this
comparison shows clearly that (annular) cluster LFs obtained at large
distances from the cluster centre are significantly affected by small
differences in the treatment of the background field star confusion and
are therefore highly uncertain.  In order to estimate the effect of
background confusion, we determined the LF slope, using identical
absolute magnitude ranges as for the cluster LFs (corrected for
incompleteness), for our two radial bins at radii $R > 72.0''$ to
investigate this effect.  The resulting LF slopes for the background LMC
stars are shown in Figs.  \ref{litcomp.fig}c and d as the dashed lines
with their associated 1-$\sigma$ uncertainties at the bottom of these
panels. 

Thus, we conclude that the value derived for the LF slope is critically
dependent on the range in (absolute) magnitude (or luminosity) used for
the fitting. However, if done consistently for the entire cluster
sample, for all radial annuli, the relative variations among LF slopes
as a function of radius are more robust.

Finally, Hunter et al.  (1997) converted their {\sl WFPC2} annular LFs
of NGC 1818 into their associated MFs, but did not find evidence for
mass segregation in this cluster for stellar masses in the range $0.85
\le m \le 9 M_\odot$.  However, they noted that the cluster core
contains brighter stars that were not included in their study due to
saturation, whereas the outer regions do not.  The exposure times of
their short-exposure F555W images are four times as long as those of our
short exposures, resulting in a significantly greater fraction of
saturated stars in their images.  Therefore, we believe that the
difference between our and their results is largely due to this
saturation effect.  Unfortunately, they did not publish background and
incompleteness corrected LFs, so that we cannot directly compare our
results.  In paper II we will discuss these differences in more detail
based on a comparison of the cluster MFs. 

\subsection{Is there a representative cluster radius?}

\begin{figure}
\psfig{figure=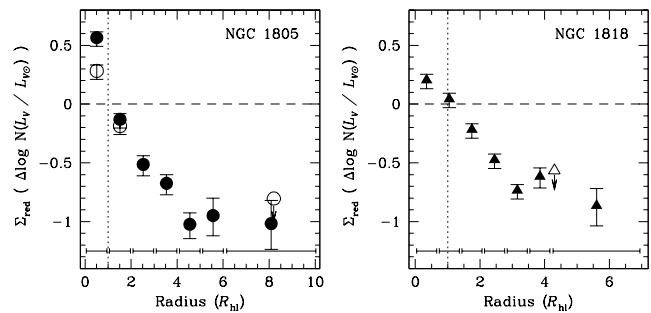,width=9cm}
\vspace*{-4.3cm}
\caption{\label{robust.fig}Deviations of the annular LFs from the global
LF as a function of radius, as discussed in the text.  The filled
symbols represent the {\sl WFPC2} data of NGC 1805 and NGC 1818,
respectively; the open symbols at large radii are their STIS
counterparts, which are upper limits due to the observed curvature of
the LFs (see text).  The error bars are dominated by statistical
fluctuations in the LFs due to poisson noise.  The horizontal bars at
the bottom of the figure indicate the radial range used to obtain the
data points; the vertical dotted lines represent the clusters'
half-light radii.}
\end{figure}

Because of the sensitivity of the LF slope to the adopted luminosity and
radial range and to the accuracy of the corrections for incompleteness
and background star contamination of single power law fits to the
annular LFs, we introduce a more robust characterisation of the presence
of luminosity segregation in the NGC 1805 and NGC 1818 in Fig. 
\ref{robust.fig}.  To minimise the sensitivity of the LF slope to these
effects, we decided to quantify the deviations of the high-luminosity
range of the annular LFs from the global LF.  All annular LFs were
normalised to the global LF in the range $-0.1 \le \log (L_V /
L_{V,\odot}) \le 1.0$; subsequently, we determined the average sum of
the differences between the global and the scaled annular LFs in the
common luminosity range $1.0 < \log (L_V / L_{V,\odot}) \le 2.57$ ($\log
(L_V / L_{V,\odot}) \simeq 2.57$ corresponds to the main sequence
turn-off), $\Sigma_{\rm red} \bigl( \Delta \log N (L_V / L_{V,\odot})
\bigr)$.  The result for the inner two radial bins of NGC 1805 are upper
limits due to the severe incompleteness effects for low-luminosity stars
in its centre (cf.  Section \ref{lumsegr.sect}).  In order to constrain
these upper limits to their most likely range, we decided to adopt the
shape of the third annulus ($14.4'' < R \le 21.6''$) matched to the
number counts of the inner annulus in the luminosity range $1.5 \le \log
L_V/L_{V,\odot} \le 1.9$ as a conservative estimate of the ``true''
inner LFs -- based on the robust assumption of no luminosity segregation
for $\log L_V/L_{V,\odot} \le 1.0$ at $R \sim 18'' = 1.53 R_{\rm hl}$ --
and repeated the above procedure for the inner two annuli.  The
resulting data points are shown as open symbols.  Since the STIS LFs do
not reach the brightest luminosity cut-off, we were forced to use a
smaller fitting range; the resulting STIS data points are therefore
upper limits, since the LFs are curved and not single power laws. 

From Fig.  \ref{robust.fig}, it is immediately clear that both clusters
are luminosity segregated to at least $R \simeq 3-4 R_{\rm hl}$, beyond
which radius the deviations become relatively constant with increasing
radius.  Although it appears that the global cluster LF is fairly well
approximated by half-light LF ($\simeq R_{\rm core}$ for these young
clusters), the LFs at these radii are not representative of the dominant
cluster stars, but are dominated by the segregated high-luminosity
stars.  In Paper III we will address this issue in more detail for our
larger LMC cluster sample, for which we will investigate whether the
cluster age affects the radius at which the local LF becomes
representative of the cluster population as a whole. 

\subsection{Dynamical evolution on short time-scales?}
\label{ejections.sect}

In Sec.  \ref{nttfields.sect} we presented evidence for a marginal
excess of bright stars ($V \le 16.0$), corresponding to the most massive
unsaturated stars in our LFs (see Paper II), surrounding NGC 1818
compared to the expected number of stars in the background field. For
NGC 1805, our results are inconclusive.

It is unlikely that they originated independently from the cluster
in the background field.  Although the LMC background field shows
significant density contrasts, including in our NTT fields, the bright
stars are not preferentially located in any of these.  In particular,
only 1 or 2 of them are located in the NGC 1818 B subcluster, while none
of them seem to be associated with the open cluster-like object at $\sim
360''$ to the North West of the cluster centre. 

Therefore, they are most likely massive stars that have been
collisionally ejected from the cluster core due to encounters with other
massive stars or binary systems.  Although standard two-body
interactions involving equal mass objects are not expected to eject a
significant number of stars over the short lifetime of NGC 1818, unequal
mass encounters, primordial mass segregation and -- in particular -- the
presence of hard binaries can have a significant effect on the dynamical
ejection of massive stars on short time-scales (e.g., Leonard \& Duncan
1988, 1990, Leonard 1995, Portegies Zwart et al.  1999, Hoogerwerf, de
Bruijne \& de Zeeuw 2000, Brandl et al.  2001).  In addition, the
dynamical relaxation time of massive stars can be significantly shorter
than that of low-mass stars (cf.  Portegies Zwart et al.  1999). 

Thus, this interpretation suggests that the cores of massive young star
clusters undergo significant dynamical evolution, even on time-scales as
short as $\sim 25$ Myr.  A more extreme example of the dynamical
evolution in the core of a very young star cluster was recently
presented for the $\lesssim 3-4$ Myr old core of 30 Dor, R136 (Brandl et
al.  2001). 

In Paper II we will explore this issue in more detail; we will convert
our LFs into the associated MFs and discuss the implications of the
observed mass segregation in the context of mass segregation at birth
versus that due to dynamical evolution. 

\section{Summary and Conclusions}

In this paper, we have presented a detailed analysis of {\sl
HST/WFPC2} and STIS imaging observations, and of supplementary
wide-field ground-based observations obtained with the NTT of two young
($\sim 10-25$ Myr) compact star clusters in the LMC, NGC 1805 and NGC
1818. 

In principle, rich compact LMC star clusters are ideal laboratories for
providing strong constraints on the universality of the IMF, in
particular because they are essentially single-age, single-metallicity
systems for which individual stars over a range of masses can easily be
resolved.  However, in order to understand the cluster IMF, a detailed
knowledge of the presence and the effects of mass segregation is
required.  Observations of various degrees of mass segregation in very
young star clusters suggest that at least some of this effect is related
to the process of star formation itself. 

In this paper, we have focussed on the analysis of the behaviour of the
stellar LF as a function of radius in these two young LMC clusters; in
Paper II, we will derive the associated MFs and discuss alternative
diagnostics to quantify mass segregation effects, such as the dependence
of cluster core radius on the adopted mass (or luminosity) range. 

Although the effects are strongest in NGC 1805, the more strongly
concentrated cluster, we present clear evidence for strong luminosity
segregation in both clusters:

\begin{enumerate}
\item The brighter cluster stars are strongly concentrated in the inner
$\sim 3-4 R_{\rm hl}$ in each cluster.  Compared to the outer cluster
regions, the fainter stars in the central annuli are significantly
underpopulated relative to the brighter stars. 

\item The LF slopes steepen with cluster radius.  In both clusters, the
LF slopes reach a stable level well beyond the clusters' half-light
radii. 

\item The brightest, saturated cluster stars ($V, I \lesssim 15.5$ and
15.0 for NGC 1805 and NGC 1818, respectively) are predominantly located
within the inner $\sim 4-8 R_{\rm hl}$.
\end{enumerate}

From a detailed analysis of the shape of the LF, we show that the
value derived for the LF slope is critically dependent on the range in
(absolute) magnitude (or luminosity) used for the fitting. 

Finally, from the wide-field NTT observations we present tentative
evidence for the presence of bright stars surrounding NGC 1818 which we
argue to be associated with the cluster.  We suggest that they are most
likely massive stars that have been collisionally ejected from the
cluster core due to unequal-mass encounters, primordial mass segregation
or collisions with (hard) binary systems.  This interpretation leads us
to suggest, therefore, that the cores of massive young stars clusters
undergo significant dynamical evolution, even on time-scales as short as
$\sim 25$ Myr. 

\section*{Acknowledgments} This paper is based on observations with the
NASA/ESA {\sl Hubble Space Telescope}, obtained at the Space Telescope
Science Institute, which is operated by the Association of Universities
for Research in Astronomy (AURA), Inc., under NASA contract NAS 5-26555. 
We thank the ESO Data Flow Operations Team for carrying out our NTT
service observations, in particular Massimo Ramella and Bruno
Leibundgut.  We also thank Dougal Mackey for his assistance. We
acknowledge insightful and constructive comments by the referee. This
research has made use of NASA's Astrophysics Data System Abstract
Service.

\end{document}

%% file: yclusparams.tex
\begin{table}
\caption[ ]{\label{sample.tab}Fundamental parameters of NGC 1805 and NGC 1818}
{\scriptsize
\begin{center}
\begin{tabular}{lllll}
\hline
\hline
& \multicolumn{1}{c}{NGC 1805} & \multicolumn{1}{c}{Ref.} &
\multicolumn{1}{c}{NGC1818} & \multicolumn{1}{c}{Ref.} \\
\hline
log( age ) [yr]            & 7.0 $\pm$ 0.05 & 1,8 & 7.2 $\pm$ 0.1 & 1,5,8 \\
                           &                &     & 7.41          & 1 \\
\mbox{[Fe/H]} (dex)        & $\sim 0.0$     & 6   & $\sim 0.0$    & 6 \\
                           &                &     & $-0.4$        & 6$^a$ \\
E$(B-V)$ (mag)             & 0.04           & 2   & 0.03          & 2 \\
                           &                &     & 0.05          & 5 \\
$(m-M)_0$ (mag)            & 18.59          & 2   & 18.58         & 2 \\
$R_{\rm core}$ (pc)$^c$    & 1.39           & 9  & 2.1 $\pm$ 0.4 & 4 \\
                           &                &     & 2.56          & 9 \\
$R_{\rm hl}$ (pc)$^c$      & 1.8            & 7   & 2.6           & 7 \\
log( $t_{\rm rh} )^b$ [yr] & $\dots$        &     & 9.0 -- 9.7    & 3 \\
Mass ($M_\odot$)           & $0.6 \times 10^4$ & 6 & $3 \times 10^4$  & 5 \\
Centre (J2000)             & RA  = 05 02 05   & 9 & RA  = 05 04 03   & 9 \\
(RA: hh mm ss;             & Dec = $-$66 06.7 & 9 & Dec = $-$66 26.0 & 9 \\
Dec: dd mm.m)              \\
\hline
\end{tabular}
\end{center}
\flushleft
{\sc Notes:} \\
$^a$ best estimate from measurements in the literature \\
$^b$ depending on the mass-to-light ratio assumed \\
$^c$ based on $(m-M)_{0,\rm LMC} = 18.5$, or $D_{\rm
LMC} = 52$ kpc. \\
{\sc References:}
1 -- Cassatella et al. (1996);
2 -- Castro et al. (2001);
3 -- Elson, Fall \& Freeman (1987a);
4 -- Elson, Freeman \& Lauer (1989);
5 -- Hunter et al. (1997);
6 -- Johnson et al. (2001);
7 -- Santiago et al. (2001);
8 -- Santos, Jr., et al. (1995);
9 -- this paper.
}
\end{table}

%% file: obslog.tex
\begin{table*}
\caption[ ]{\label{obslog.tab}Overview of the observations
}
{\scriptsize
\begin{center}
\begin{tabular}{llllrllrc}
\hline
\hline
\multicolumn{1}{c}{Object} & \multicolumn{1}{c}{Field} &
\multicolumn{1}{c}{Detector} & \multicolumn{1}{c}{Filter} &
\multicolumn{1}{c}{Exposure} & \multicolumn{1}{c}{RA$^a$} & 
\multicolumn{1}{c}{Dec$^a$} & \multicolumn{1}{c}{Position} &
\multicolumn{1}{c}{Date (UT)} \\ 
& & & & time (s) & \multicolumn{2}{c}{(J2000)} & angle ($^\circ$)$^b$ &
(dd/mm/yyyy) \\
\hline
NGC 1805 & CEN  & {\sl WFPC2} & F555W & 3x5    & 05:02:21.652 & $-$66:06:43.110 &  $-$89.88 & 25/07/1998 \\
         &      &             &       & 3x140  & \\
         &      &             & F814W & 3x20   & & & & 25/07/1998 \\
         &      &             &       & 3x300  & \\
         & HALF &             & F555W & 2x800  & 05:02:24.533 & $-$66:06:12.752 & $-$177.09 & 29/04/1998 \\
         &      &             &       & 900    & \\
         &      &             & F814W & 3x800  & & & & 28/04/1998 \\
         &      &             &       & 900    & \\
         &      & STIS     & F28x50LP & 5x2950 & 05:02:24.668 & $-$66:06:14.190 &     53.06 & 13/03/1998 \\
         & Wide & NTT         & {\it V}$_{606}$\strut$^c$ & 3x60 & 05:04:13.9 & $-$66:26:05.5 & 88.50 & 10/01/2000 \\
         &      &             &       & 2x20   & \\
         &      &             & {\it I}$_{610}$\strut$^c$ & 3x60 & 05:04:13.9 & $-$66:26:05.5 & 88.50 & 10/01/2000 \\
         &      &             &       & 2x10   & \\
NGC 1818 & CEN  & {\sl WFPC2} & F555W & 3x5    & 05:04:14.135 & $-$66:26:05.647 &  $-$45.09 & 25/09/1998 \\
         &      &             &       & 3x140  & \\
         &      &             & F814W & 3x20   & & & & 25/09/1998 \\
         &      &             &       & 3x300  & \\
         & HALF &             & F555W & 2x800  & 05:04:12.261 & $-$66:26:33.152 & $-$177.09 & 30/04/1998 \\
         &      &             &       & 900    & \\
         &      &             & F814W & 3x800  & & & & 30/04/1998 \\
         &      &             &       & 900    & \\
         &      & STIS     & F28x50LP & 5x2950 & 05:04:12.345 & $-$66:26:34.081 &    167.06 & 29/07/1998 \\
         & Wide & NTT         & {\it V}$_{606}$\strut$^c$ & 4x60 & 05:04:13.9 & $-$66:26:05.5 & 88.50 & 10/01/2000 \\
         &      &             &       & 2x20   & \\
         &      &             & {\it I}$_{610}$\strut$^c$ & 3x60 & 05:04:13.9 & $-$66:26:05.5 & 88.50 & 10/01/2000 \\
         &      &             &       & 2x10   & \\
\hline
\multicolumn{9}{c}{\sc Background fields}\\
NGC 1805-par &  & {\sl WFPC2} & F555W & 2x600  & 05:01:28.193 & $-$66:01:04.943 &     44.72 & 08/12/1997 \\
         &      &             & F814W & 2x400  & & & & 08/12/1997 \\
         &      & STIS     & F28x50LP & 2x1350 & 05:02:22.939 & $-$65:57:07.421 &  $-$34.94 & 08/12/1997 \\
NGC 1818-par &  & {\sl WFPC2} & F555W & 2x600  & 05:03:24.113 & $-$66:20:04.149 &     48.81 & 11/12/1997 \\
         &      &             & F814W & 2x400  & & & & 11/12/1997 \\
         &      & STIS     & F28x50LP & 2x1350 & 05:03:18.941 & $-$66:19:12.111 &  $-$69.73 & 05/12/1997 \\
Background-1 &  & {\sl WFPC2} & F555W & 6x1300 & 05:11:27.853 & $-$65:29:01.693 &  $-$72.00 & 13/08/1998 \\
         &      &             & F814W & 4x1300 & & & & 26/08/1998 \\
\hline
\end{tabular}
\end{center}
\flushleft
$^a$ For {\sl WFPC2}, centre of the PC; for STIS observations, field
centre; for NTT observations, telescope pointing; $^b$ East w.r.t. 
North; $^c$ These are filter names in the NTT nomenclature, and not
related to {\sl HST} filter names.
}
\end{table*}

%% file: sources.tex
\begin{table*}
\caption[ ]{\label{sources.tab}Source detections and saturation levels
}
{\scriptsize
\begin{center}
\begin{tabular}{ccccrrc}
\hline
\hline
\multicolumn{1}{c}{Object} & \multicolumn{1}{c}{Field} &
\multicolumn{1}{c}{Instrument} & \multicolumn{1}{c}{Filter} &
\multicolumn{1}{c}{\# Stars} & \multicolumn{1}{c}{\# Rej.$^a$} &
\multicolumn{1}{c}{Saturation} \\
& & & & \multicolumn{1}{c}{(total)} & & \multicolumn{1}{c}{level
(mag)$^b$}\\ 
\hline
NGC 1805 & CEN  & {\sl WFPC2} & F555W & 5457 &  10 & 15.5, 18.6 \\
         &      &             & F814W & 5457 &  62 & 15.5, 18.6 \\
         & HALF &             & F555W & 5850 & 173 & 20.6 \\
         &      &             & F814W & 5850 & 127 & 20.5 \\
         & All  &             &       &10202 \\
         & HALF & STIS     & F28x50LP &  612 &  16 & 18.0 \\
         & Wide & NTT     & $V_{606}$ & 2000 &  35 & 15.5 \\
         &      &         & $I_{610}$ & 2000 &  21 & 15.6 \\
NGC 1818 & CEN  & {\sl WFPC2} & F555W & 6881 &  25 & 15.0, 18.3 \\
         &      &             & F814W & 6881 &  62 & 15.0, 18.4 \\
         & HALF &             & F555W & 7368 & 823 & 20.5 \\
         &      &             & F814W & 7368 & 606 & 20.4 \\
         & All  &             &       &12833 \\
         & HALF & STIS     & F28x50LP &  896 &   5 & 18.0 \\
         & Wide & NTT     & $V_{606}$ & 2371 & 117 & 15.5 \\
         &      &         & $I_{610}$ & 2371 & 108 & 15.6 \\
\hline
\multicolumn{7}{c}{\sc Background fields}\\
NGC 1805-par &  & {\sl WFPC2} & F555W &      &     & 20.3 \\
             &  &             & F814W &      &     & 20.0 \\
NGC 1818-par &  &             & F555W &      &     & 20.3 \\
             &  &             & F814W &      &     & 20.0 \\
Background-1 &  &             & F555W &      &     & 22.0 \\
             &  &             & F814W &      &     & 21.5 \\
\hline
\end{tabular}
\end{center}
\flushleft
{\sc Notes:} $^a$ This refers to the number of stars rejected from the
final sources lists due to abnormally high background levels, or steep
background gradients due to the vicinity of saturated stars; CEN: long
exposures only; $^b$ CEN exposures: short, long. 
}
\end{table*}

%% file: apcor.tex
\begin{table}
\caption[ ]{\label{apcor.tab}Aperture corrections to $0.5''$ apertures.
}
{\scriptsize
\begin{center}
\begin{tabular}{ccllll}
\hline
\hline
\multicolumn{1}{c}{Object} & \multicolumn{1}{c}{Filter} &
\multicolumn{1}{c}{Chip} & \multicolumn{1}{c}{$a$} & 
\multicolumn{2}{c}{$b$} \\ 
\hline
NGC 1805 & F555W & PC  & 0.3174 & 1.13   & $\times 10^{-4}$ \\
         &       & WF2 & 0.1511 & 7.6783 & $\times 10^{-5}$ \\
         &       & WF3 & 0.1811 & 6.6327 & $\times 10^{-5}$ \\
         &       & WF4 & 0.1488 & 1.0811 & $\times 10^{-4}$ \\
         & F814W & PC  & 0.5071 & 1.95   & $\times 10^{-4}$ \\
         &       & WF2 & 0.1797 & 8.0328 & $\times 10^{-5}$ \\
         &       & WF3 & 0.2116 & 3.0135 & $\times 10^{-5}$ \\
         &       & WF4 & 0.1829 & 8.7969 & $\times 10^{-5}$ \\
         & F28x50LP & STIS & 0.468 \\
NGC 1818 & F555W & PC  & 0.3192 & 1.098  & $\times 10^{-4}$ \\
         &       & WF2 & 0.1276 & 1.3832 & $\times 10^{-4}$ \\
         &       & WF3 & 0.2020 & 9.9948 & $\times 10^{-6}$ \\
         &       & WF4 & 0.1424 & 1.3596 & $\times 10^{-4}$ \\
         & F814W & PC  & 0.5469 & 8.9871 & $\times 10^{-5}$ \\
         &       & WF2 & 0.1959 & 9.8792 & $\times 10^{-5}$ \\
         &       & WF3 & 0.2562 & 2.2080 & $\times 10^{-5}$ \\
         &       & WF4 & 0.2088 & 1.0289 & $\times 10^{-4}$ \\
         & F28x50LP & STIS & 0.463 \\
\hline
\end{tabular}
\end{center}
}
\end{table}

%% file: offsets.tex
\begin{table}
\caption[ ]{\label{offsets.tab}Photometric offsets applied to match the
photometry of exposures with different exposure times.
}
{\scriptsize
\begin{center}
\begin{tabular}{cclrr}
\hline
\hline
\multicolumn{1}{c}{Object} & \multicolumn{1}{c}{Filter} &
\multicolumn{1}{c}{Chip} & \multicolumn{1}{c}{mag(long) --} & 
\multicolumn{1}{c}{mag(CEN) --} \\ 
& & & \multicolumn{1}{c}{mag(short)} & \multicolumn{1}{c}{mag(HALF)} \\
\hline
NGC 1805 & F555W & PC  &  0.072 $\pm$ 0.074 \\
         &       & WF2 &  0.031 $\pm$ 0.099 \\
         &       & WF3 &  0.044 $\pm$ 0.111 \\
         &       & WF4 &  0.050 $\pm$ 0.124 \\
         &       & All &                    & $-$0.027 $\pm$ 0.053 \\
         & F814W & PC  &  0.064 $\pm$ 0.047 \\
         &       & WF2 & $-$0.011 $\pm$ 0.042 \\
         &       & WF3 & $-$0.009 $\pm$ 0.053 \\
         &       & WF4 &  0.004 $\pm$ 0.080 \\
         &       & All &                    & $-$0.094 $\pm$ 0.063 \\
NGC 1818 & F555W & PC  &  0.069 $\pm$ 0.060 \\
         &       & WF2 &  0.041 $\pm$ 0.065 \\
         &       & WF3 &  0.030 $\pm$ 0.048 \\
         &       & WF4 &  0.041 $\pm$ 0.069 \\
         &       & All &                    &  0.016 $\pm$ 0.092 \\
         & F814W & PC  &  0.052 $\pm$ 0.053 \\
         &       & WF2 &  0.028 $\pm$ 0.050 \\
         &       & WF3 &  0.017 $\pm$ 0.050 \\
         &       & WF4 &  0.024 $\pm$ 0.056 \\
         &       & All &                    &  0.063 $\pm$ 0.150 \\
\hline
\end{tabular}
\end{center}
}
\end{table}

%% file: lmcrev1.bbl
\begin{thebibliography}{}

\bibitem[]{} Andreuzzi G., De Marchi G., Ferraro F.R., Paresce F.,
Pulone L., Buonanno R., 2001, A\&A, 372, 851

\bibitem[]{} Beaulieu S.F., Elson R.A.W., Gilmore G.F., Johnson R.A.,
Tanvir N., Santiago B.X., 1999, in: New Views of the Magellanic Clouds,
IAU Symp. 190, Chu Y.-H., Suntzeff N., Hesser J., Bohlender D., eds.,
Victoria, Canada, p. 460

\bibitem[]{} Beaulieu S.F., Gilmore G.F., Elson R.A.W., Johnson R.A.,
Santiago B.X., Sigurdsson S., Tanvir N., 2001, AJ, 121, 2816

\bibitem[]{} Brandl B., Chernoff D.F., Moffat A.F.J., 2001, in:
``Extragalactic Star Clusters'', IAU Symposium 207, Puc\'on (Chile), March
2001, eds.  Grebel E.K., Geisler D., in press

\bibitem[]{} Brandl B., Sams B.J., Bertoldi F., Eckart A., Genzel R.,
Drapatz S., Hofmann R., L\"owe M., Quirrenbach A., 1996, ApJ, 466, 254

\bibitem[]{} Caloi V., Cassatella A., 1998, A\&A, 330, 492

\bibitem[]{} Campbell B., et al., 1992, AJ, 104, 1721

\bibitem[]{} Carpenter J.M., Meyer M.R., Dougados C., Strom S.E.,
Hillenbrand L.A., 1997, AJ, 114, 198

\bibitem[]{} Cassatella A., Barbero J., Brocato E., Castellani V., Geyer
E.H., 1996, A\&A, 306, 125

\bibitem[]{} Castro R., Santiago B.X., Gilmore G.F., Beaulieu S.,
Johnson R.A., 2001, MNRAS, 326, 333

\bibitem[]{} Da Costa G.S., 1982, AJ, 87, 990

\bibitem[]{} de Grijs R., Gilmore G.F., Johnson R.A., Mackey A.D., 2001,
MNRAS, submitted (Paper II)

\bibitem[]{} de Grijs R., O'Connell R.W., Gallagher J.S., 2001, AJ, 121,
768 

\bibitem[]{} Elson R.A.W., Fall S.M., 1988, AJ, 96, 1383

\bibitem[]{} Elson R.A.W., Fall S.M., Freeman K.C., 1987, ApJ, 323, 54

\bibitem[]{} Elson R.A.W., Freeman K.C., Lauer T.R., 1989, ApJ, 347, 69

\bibitem[]{} Elson R.A.W., Sigurdsson S., Davies M.B., Hurley J.,
Gilmore G.F., 1998, MNRAS, 300, 857

\bibitem[]{} Elson R.A.W., Tanvir N., Gilmore G.F., Johnson R.A.,
Beaulieu S.F., 1999, in: New Views of the Magellanic Clouds,
IAU Symp. 190, Chu Y.-H., Suntzeff N., Hesser J., Bohlender D., eds.,
Victoria, Canada, p. 417

\bibitem[]{} Fischer P., Pryor C., Murray S., Mateo M., Richtler T.,
1998, AJ, 115, 592

\bibitem[]{} Grebel E.K., 1997, A\&A, 317, 448

\bibitem[]{} Harris W.E., Canterna R., 1980, ApJ, 239, 815

\bibitem[]{} Hillenbrand L.A., 1997, AJ, 113, 1733

\bibitem[]{} Hillenbrand L.A., Carpenter J.M., 2000, ApJ, 540, 236

\bibitem[]{} Hillenbrand L.A., Hartmann L.E., 1998, ApJ, 492, 540

\bibitem[]{} Holtzman J.A., et al., 1995a, PASP, 107, 156

\bibitem[]{} Holtzman J.A., Burrows C.J., Casertano S., Hester J.J.,
Trauger J.T., Watson A.M., Worthey G., 1995b, PASP, 107, 1065

\bibitem[]{} Hoogerwerf R., de Bruijne J.H.J., de Zeeuw P.T., 2000, ApJ,
544, L133

\bibitem[]{} Houdashelt M.L., Wyse R.F.G., Gilmore G.F., 2001, PASP,
113, 49

\bibitem[]{} Hunter D.A., Light R.M., Holtzman J.A., Lynds R., O'Neil
E.J., Grillmair C.J., 1997, ApJ, 478, 124

\bibitem[]{} Hunter D.A., Shaya E.J., Holtzman J.A., Light R.M., O'Neil
E.J., Lynds R., 1995, ApJ, 448, 179

\bibitem[]{} Johnson R.A., Beaulieu S.F., Gilmore G.F., Hurley J.,
Santiago B.X., Tanvir N.R., Elson R.A.W., 2001, MNRAS, 324, 367

\bibitem[]{} Kontizas M., Hatzidimitriou D., Bellas-Velidis I.,
Gouliermis D., Kontizas E., Cannon R.D., 1998, A\&A, 336, 503

\bibitem[]{} Krist J., Hook R., 1997, The Tiny Tim User's Guide, version
4.4

\bibitem[]{} Lada E.A., DePoy D.L., Evans N.J., Gatley I., 1991, ApJ,
371, 171

\bibitem[]{} Leonard P.J.T., 1995, MNRAS, 277, 1080

\bibitem[]{} Leonard P.J.T., Duncan M.J., 1988, AJ, 96, 222

\bibitem[]{} Leonard P.J.T., Duncan M.J., 1990, AJ, 99, 608

\bibitem[]{} Larson R.B., 1993, in: The Globular Cluster--Galaxy
Connection, Smith G.H., Brodie J.P., eds., ASP Conf. Ser. 48, San
Francisco: ASP, p. 675

\bibitem[]{} Malumuth E.M., Heap S.R., 1994, AJ, 107, 1054

\bibitem[]{} Mateo M., Hodge P., 1986, ApJS, 60, 893

\bibitem[]{} Mateo M., Hodge P., 1987, ApJ, 320, 626

\bibitem[]{} Mateo M., Hodge P., Schommer R.A., 1986, ApJ, 311, 113

\bibitem[]{} Nemec J.M., Harris H.C., 1987, ApJ, 316, 172

\bibitem[]{} Papenhausen P., Schommer R.A., 1988, in: The Harlow
Shapley Symposium On Globular Cluster Systems in Galaxies, Grindlay J.,
Philip A.G.D., eds., IAU Symp. 126, Dordrecht: Kluwer, p. 565

\bibitem[]{} Paresce F., De Marchi G., Jedrzejewski R., 1995, ApJ, 442,
L57

\bibitem[]{} Portegies Zwart S.F., Makino J., McMillan S.L.W., Hut P.,
1999, A\&A, 348, 117

\bibitem[]{} Ratnatunga K.U., Bahcall J.N., 1985, ApJS, 59, 63

\bibitem[]{} Rieke G.H., Lebofsky M.J., 1985, ApJ, 288, 618

\bibitem[]{} Santiago B.X., Beaulieu S., Johnson R., Gilmore G.F., 2001,
A\&A, 369, 74 (SBJG)

\bibitem[]{} Santos Jr.  J.F.C., Bica E., Clar\'\i a J.J., Piatti A.E.,
Girardi L.A., Dottori H., 1995, MNRAS, 276, 1155

\bibitem[]{} Saviane I., Piotto G., Fagotto F., Zaggia S., Capaccioli
M., Aparicio A., 1998, A\&A, 333, 479

\bibitem[]{} Shara M.M., Drissen L., Bergeron L.E., Paresce F., 1995,
ApJ, 441, 617

\bibitem[]{} Sirianni M., Nota A., De Marchi G., Leitherer C., Clampin
M., 2001, in: ``Extragalactic Star Clusters'', IAU Symposium 207, Puc\'on
(Chile), March 2001, eds.  Grebel E.K., Geisler D., in press

\bibitem[]{} Stetson P.B., 1987, PASP, 99, 91

\bibitem[]{} Subramaniam A., Sagar R., Bhatt H.C., 1993, A\&A, 273, 100

\bibitem[]{} Testi L., Palla F., Prusti T., Natta A., Maltagliati S.,
1997, A\&A, 320, 159

\bibitem[]{} Westerlund B.E., 1961, Uppsala Astr. Obs. Ann., 5(1)

\bibitem[]{} Whitmore B., Heyer I., Casertano S., 1999, PASP, 111, 1559

\bibitem[]{} Will J.-M., Bomans D.J., Tucholke H.-J., de Boer K.S.,
Grebel E.K., Richtler T., Seggewiss W., Vallenari A., 1995, A\&AS, 112,
367

\end{thebibliography}
